\documentclass[aps,prl,reprint]{revtex4-2}

\usepackage{graphicx}
\usepackage{amsmath,amssymb} 
\usepackage{physics}
\graphicspath{figures/}
\usepackage{dcolumn}
\usepackage{amssymb}
\usepackage{amsmath}
\usepackage{fancyhdr}
\usepackage{bm}
\usepackage{times}
\usepackage[colorlinks,linkcolor=blue,anchorcolor=blue,citecolor=blue,urlcolor=blue]{hyperref}
\usepackage{bibunits}

\begin{document}

\title{Three-body interaction in a magnon-Andreev-superconducting qubit system: collapse-revival phenomena and entanglement redistribution}
\author{Sheng Zhao}
\affiliation{Ministry of Education Key Laboratory for Nonequilibrium Synthesis and Modulation of Condensed Matter, Shaanxi Province Key Laboratory of Quantum Information and Quantum Optoelectronic Devices, School of Physics, Xi'an Jiaotong University, Xi'an 710049, China}
\author{Peng-Bo Li}
\email{lipengbo@mail.xjtu.edu.cn}
\affiliation{Ministry of Education Key Laboratory for Nonequilibrium Synthesis and Modulation of Condensed Matter, Shaanxi Province Key Laboratory of Quantum Information and Quantum Optoelectronic Devices, School of Physics, Xi'an Jiaotong University, Xi'an 710049, China}

\date{\today}

\begin{abstract}
Three-body interactions are fundamental for realizing novel quantum phenomena beyond pairwise physics, yet their implementation—particularly among distinct quantum systems—remains challenging. Here, we propose a hybrid quantum architecture comprising a magnonic mode (in a YIG sphere), an Andreev spin qubit (ASQ), and a superconducting qubit (SCQ), to realize a strong three-body interaction at the single-quantum level. Leveraging the spin-dependent supercurrent and circuit-integration flexibility of the ASQ, it is possible to engineer a strong tripartite coupling that jointly excites both qubits upon magnon annihilation (or excites magnons and SCQs upon ASQ deexcitation). Through analytical and numerical studies, we demonstrate that this interaction induces synchronized collapse and revival in qubit populations when the magnon is initially prepared in a coherent state. Notably, during the collapse region—where populations remain static—the entanglement structure undergoes a dramatic and continuous
reorganization. We show that the genuine tripartite entanglement is redistributed into bipartite entanglement between the two qubits, and vice versa, with the total entanglement conserved. These phenomena, unattainable via two-body couplings, underscore the potential of three-body interactions for exploring intrinsically new quantum effects and advancing hybrid quantum information platforms.

\end{abstract}
\maketitle
\begin{bibunit}[apsrev4-2]
\textit{Introduction.}---
Coherent interactions between different degrees of freedom in hybrid quantum systems are essential for quantum science and technology \cite{RevModPhys.85.623}. Although two-body interactions form the basis for most natural phenomena \cite{frisk2019ultrastrong}, three-body interactions play a pivotal role in various physical contexts: the physics of nuclei~\cite{loiseau1967three}, quantum blockade~\cite{chakram2022multimode}, quantum error correction~\cite{KITAEV20062}, and holography~\cite{PhysRevLett.70.3339,PhysRevD.94.106002}. Importantly, three-body interactions can give rise to novel physical phenomena that are inaccessible with only two-body interactions. Therefore, the realization of three-body interactions is of fundamental importance. However, it is a challenge to achieve the three-body interaction within the strong coupling regime between a bosonic mode and two different qubits in  hybrid quantum setups.

 Recently, Andreev (or superconducting) spin qubits (ASQ) have garnered significant interest as a promising platform for quantum information processing. The qubit combines the benefits of both superconducting qubits (SCQ) and spin qubits to encode its state in the spin degree of freedom of a quasiparticle localized in a semiconducting quantum dot junction~\cite{PhysRevLett.90.226806,PhysRevB.81.144519,doi:10.1126/science.abf0345,PhysRevLett.131.097001,pita2023direct,pita2024strong}. It also has the advantages of long lifetimes and small size~\cite{PhysRevLett.131.097001,pita2023direct}, making it an attractive candidate for large-scale quantum devices~\cite{PRXQuantum.6.010308}.  Recent experimental realizations have been achieved for the coherent manipulation of a single ASQ~\cite{doi:10.1126/science.abf0345, PhysRevLett.131.097001}, the supercurrent-mediated coupling between two distant ASQs~\cite{pita2024strong}, and the photon-mediated coupling between two ASQs~\cite{cheung2024photon}. Due to spin-orbit coupling, the quantum dot junction has a spin-dependent supercurrent~\cite{PhysRevB.81.144519,PhysRevLett.131.097001}, which allows it to be embedded in superconducting circuits as a component. This characteristic enables ASQs to be easily coupled with other components in superconducting circuits~\cite{pita2023direct}. So, through this outstanding feature of ASQs, it is possible and appealing to realize strong three-body interactions in a hybrid quantum architecture.

Quantum magnonics~\cite{chumak2015magnon,tabuchi2016quantum,Lachance-Quirion_2019,LI20211,YUAN20221,pirro2021advances,zhang2015magnon,PhysRevB.92.184419,PhysRevLett.121.087205,PhysRevA.99.021801,PhysRevApplied.12.034001,YU20211,PhysRevB.108.L180409,Yang_2024,PhysRevLett.132.193601,PhysRevLett.133.043601,PhysRevA.111.033712,8jy6-fp5x,shen2025cavity} is also a rapidly developing field, which provides a promising platform for studying fundamental quantum phenomena and their applications. Magnon is the quanta of collective spin excitations of magnetic materials [like yttrium iron garnet (YIG)]~\cite{barker2016thermal,collet2016generation,princep2017full,wei2022giant,zhang2016cavity}. Owing to their  widely tunable frequency and strong coupling to various quantum systems, magnons also serve as a promising carrier for quantum information. Magnon-photon coupling~\cite{PhysRevLett.104.077202,zhang2015cavity,PhysRevB.96.094412,PhysRevB.99.134426,PhysRevLett.132.116701}, magnon-phonon coupling~\cite{doi:10.1126/sciadv.1501286,PhysRevLett.128.013602,PhysRevLett.129.243601,PhysRevLett.129.243601,PhysRevLett.125.147201,10.1063/5.0047054}, magnon-solid-state spin coupling~\cite{PhysRevLett.130.073602,PhysRevLett.125.247702,PhysRevA.103.043706,PhysRevB.105.245310,andrich2017long,PRXQuantum.2.040314,PhysRevApplied.16.064008,PhysRevB.105.075410,doi:10.1126/sciadv.adi2042} and magnon-SCQ coupling~\cite{PhysRevLett.129.037205,PhysRevLett.125.117701} have been proposed and analyzed. However, previous studies mostly focus on coherent pairwise interactions between magnons and other completely
different physical systems; it seems to us that a genuine three-body interaction among magnons, ASQs and SCQs  to construct hybrid quantum setups remains unexplored.

In this work, we propose and investigate three-body interactions in a hybrid quantum system consisting of a YIG sphere, an ASQ, and a SCQ.
We perform both analytic derivations  and  numerical simulations to verify our central results.
The ASQ possesses spin–supercurrent coupling that enables it to easily couple with the SCQ. Utilizing this property and considering the influence of the magnetic flux induced by the YIG sphere, it is feasible to realize strong three-body interaction at the single quantum level among the magnon, the SCQ and the ASQ.  This kind of three-body interaction, which involves a bosonic mode and two solid-state qubits,  has no counterpart in previous studies. It enables the joint excitations of magnons and SCQs upon ASQ deexcitation (or excites two solid-state qubits upon magnon annihilation), allowing for simultaneous manipulation of two solid-state qubits via magnons.

Furthermore, by preparing the magnon in a coherent state, with this three-body interaction, we show that the population oscillations of the two qubits exhibit synchronized collapse and revival. Most notably, during the collapse region—where the populations remain stationary—the entanglement continues to evolve dynamically. Specifically, we confirm that the entanglement between the magnon and the rest of the system consists solely of residual entanglement, which first increases and then decreases, with no bipartite entanglement ever forming between the magnon and each qubit. Moreover, the entanglement between each qubit and the rest of the system also displays collapse-revival behavior, remaining constant during the collapse region. This composite entanglement comprises both bipartite qubit-qubit entanglement and the minimal residual entanglement. These two components exhibit a trade-off in the collapse region: an increase in one necessarily leads to a decrease in the other. This leads to a remarkable conclusion: during the collapse region, although the energy remains constant, the genuine tripartite entanglement is dynamically redistributed into bipartite entanglement between the two qubits, and vice versa, with the total entanglement conserved. These phenomena have never been uncovered before and cannot be achieved through two-body interactions, highlighting the unique potential of three-body interactions in revealing genuinely novel quantum effects.

\begin{figure}[t]
    \centering
    \includegraphics[width=0.5\textwidth]{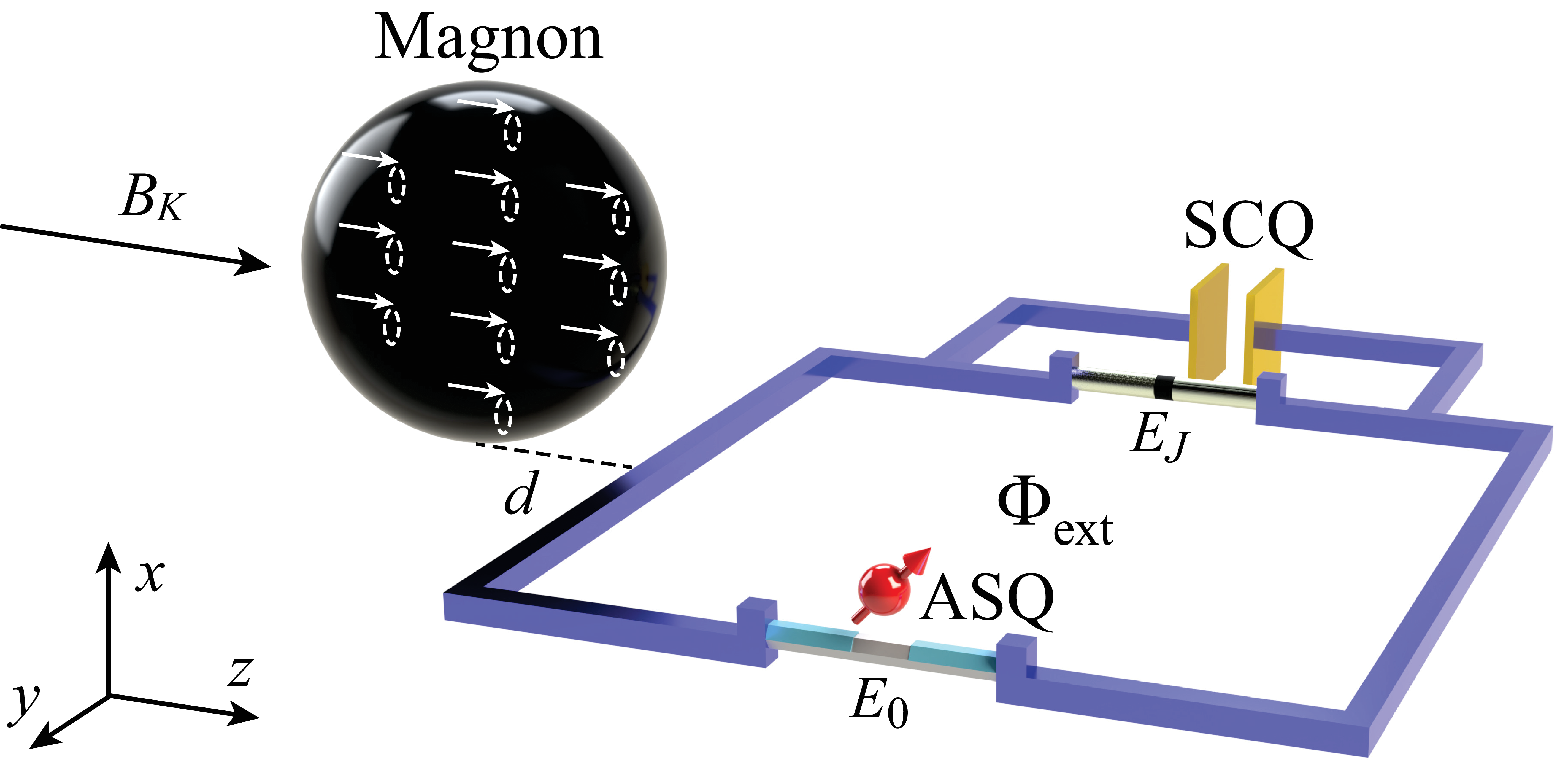}
    \caption{The hybrid quantum system consisting of the magnon, ASQ from the quantum dot junction, and the SCQ from the SQUID and capacitor.}
    \label{FIG1}
\end{figure}

\textit{The model.}---We investigate a hybrid quantum system consisting of a YIG sphere and a superconducting circuit constructed with a quantum dot junction, a conventional Josephson junction and a capacitor, as depicted in Fig.~\ref{FIG1}. The ASQ is implemented via the quantum dot junction. The Hamiltonian of ASQ includes the spin-dependent energy $E_\text{SO}$ due to spin-orbit coupling and the Zeeman energy arising from the interaction between spin and an external magnetic field $B_Z$ applied to the junction~\cite{PhysRevLett.131.097001,pita2023direct,pita2024strong}, which is given by
\begin{equation}
    H_\mathrm{ASQ}= -E_\text{SO}\boldsymbol{\sigma}\cdot\boldsymbol{n}\sin{\varphi_1}+\frac{1}{2}\boldsymbol{E}_\text{Z} \cdot\boldsymbol{\sigma},
    \label{EQ1}
\end{equation}
where $\boldsymbol{\sigma}$ is the spin operator, $\boldsymbol{n}$ is the unit vector along the spin-polarization direction, $\varphi_1$ is the phase difference across the quantum dot junction and $\boldsymbol{E}_\text{Z}$ is the Zeeman field.  In the limit of $\vert\boldsymbol{E}_\text{Z}\vert\gg E_\text{SO}$, the free Hamiltonian of the ASQ is mainly dominated by the second term, while the first term gives rise to interactions with other subsystems.
The free Hamiltonian of ASQ can be simplified to $\hat H_\text{ASQ}=\hbar\omega_a\hat\sigma^z_a/2$, where $\omega_a\simeq \vert\boldsymbol{E}_\text{Z}\vert/\hbar$ is the frequency of ASQ~\cite{pita2023direct,PRXQuantum.6.010308,SM} and $\sigma^z_a=\ket{\uparrow}\bra{\uparrow}-\ket{\downarrow }\bra{\downarrow }$ is the Pauli $z$ operator with the eigenstates of ASQ $\{\bra{\downarrow },\ket{\uparrow }\}$.

Furthermore, the quantum dot junction and conventional Josephson junction, with Josephson energies $E_0$ and $E_{J}$ respectively, form a superconducting quantum interference device (SQUID) with an external flux $\Phi_\mathrm{ext}$ through it. The SCQ is obtained by connecting the SQUID loop in parallel with the capacitor that has charging energy $E_C$. Its Hamiltonian reads $ \hat{H}_\mathrm{SCQ}=4E_{C}\hat{N}^{2} -E_{J}^\text{sum}S(\phi_\text{ext})\cos{( \hat{\varphi}-\varphi_0 )}$ with charge number $\hat{N}$ and average phase difference $\hat{\varphi}$, where $E_{J}^\text{sum}=E_J+E_{0}$, ${S(\phi_\text{ext})=\sqrt{\cos^{2}{\phi_\text{ext}}+a^{2}\sin^{2}{\phi_\text{ext}}}}$ with junction asymmetry $a=(E_J-E_0)/E_{J}^{\mathrm{sum}}$ and phase $\phi_\text{ext}=\pi\Phi_\text{ext}/\Phi_0$, and $\varphi_0=\arctan(a\tan{\phi_\text{ext}})$.
In the two-level subspace $\{\ket{g},\ket{e}\}$ of the SCQ, the Hamiltonian of SCQ is rewritten as $\hat{H}_\mathrm{SCQ}=\frac{\hbar\omega_s}{2} \hat{\sigma}_s^z$ in the transmon regime~\cite{SM,PhysRevA.76.042319,RevModPhys.93.025005}. Here $\hbar\omega_s=\sqrt{8 E_C E_J^\text{sum} S(\phi_\mathrm{ext})} - E_C$ is the qubit excitation energy.

The YIG sphere with radius $R_K$ is placed at a distance of $d$ from the circuit in the $z$ direction and $R_K$ in the $x$ direction. Applying a uniform bias magnetic field $B_K$ to the sphere, long-lived spin wave mode can be excited. In the system, only the Kittel mode is considered, where all spins precess in phase and have the identical amplitude~\cite{tabuchi2016quantum}. The free Hamiltonian of the Kittel mode is expressed as $\hat{H}_\mathrm{M}=\hbar\omega_m\hat{m}^{\dagger}\hat{m}$, where $\hat{m}$ ($\hat{m}^{\dagger}$) represents the annihilation (creation) operator of magnon and $\omega_m=\gamma_0B_K $ is the resonance frequency with the gyromagnetic ratio $\gamma_0$~\cite{ZARERAMESHTI20221}.

The magnetic moment ${\boldsymbol{\mu}}$ of the YIG sphere emits a stray field described by $ \boldsymbol{B}({\boldsymbol{\mu}})={\mu_0}/(4\pi r^3)[{3\boldsymbol{r}(\boldsymbol{\mu}\cdot\boldsymbol{r})}/{r^2}-\boldsymbol{\mu}]$~\cite{PhysRevLett.124.163604}, where $\mu_0$ is the permeability of the free space, and $\boldsymbol{r}$ is the position vector from the center of the YIG sphere to any location on the SQUID loop.  By the quantized magnetic moment $\hat{\mu}_x=\hbar\gamma_0\sqrt{N_S/2}(\hat{m}^\dagger+\hat{m})$, $\hat{\mu}_y=i\hbar\gamma_0\sqrt{N_S/2}(\hat{m}-\hat{m}^\dagger)$ and $\hat{\mu}_z=\hbar\gamma_0\hat{m}^\dagger\hat{m}$, with the total number of spins $N_S$, the magnetic field can be quantized as $\hat{\boldsymbol{B}}=(\hat{B}_x,\hat{B}_y,\hat{B}_z)$. The magnetic field $\hat{\boldsymbol{B}}$ induces a flux through the SQUID loop, that reads
\begin{equation}
    \Phi(\hat{\boldsymbol{\mu}})=\iint\boldsymbol{B}(\hat{\boldsymbol{\mu}})\cdot d\boldsymbol{S}=\Phi_\text{YIG}(\hat{m}+\hat{m}^\dagger),
    \label{Eq 2}
\end{equation}
where $\Phi_\text{YIG}$ is related to the relative positions and relative sizes of the YIG sphere and the SQUID loop~\cite{SM}.

\begin{figure*}[htbp]
    \centering
    \includegraphics[width=0.9\textwidth]{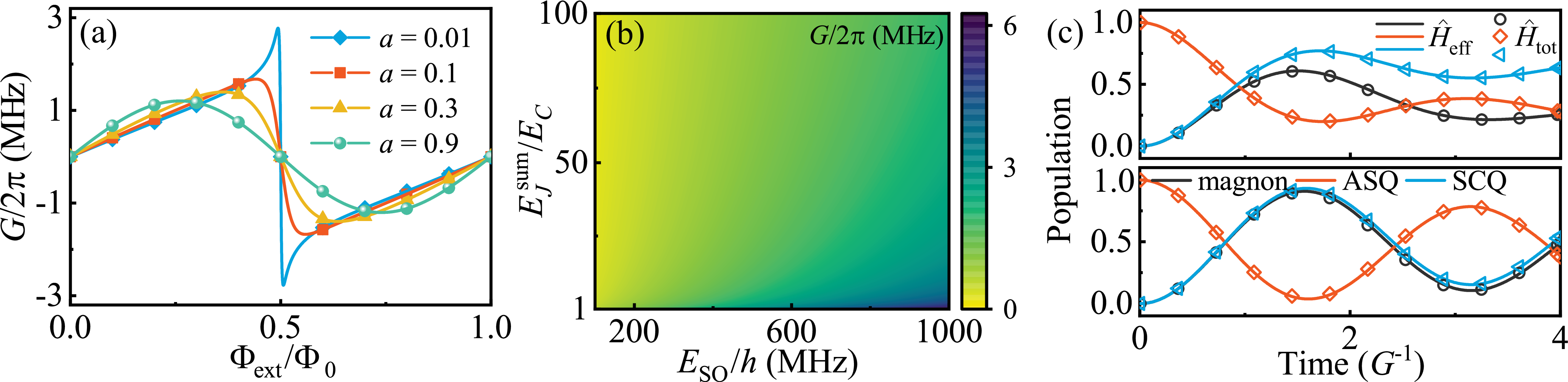}
    \caption{(a) Variation of the coupling strengths $G$ with the external flux $\Phi_\mathrm{ext}$ for different asymmetry $a$. (b) Contour maps of the coupling strengths $G$ versus the ratio $E_{J}^{\mathrm{sum}}/E_C$ and spin-dependent energy $E_\text{SO}$ with $a=0.1$. (c) The three-body interaction induce dynamical evolution under the resonance condition $\omega_m=\omega_a+\omega_s$. In the upper panel, the dissipation rates are chosen as those in the main text. In the lower panel, the dissipation rates are reduced as $\kappa_m/2\pi=0.1~\mathrm{MHz}$ and $\gamma_a/2\pi=0.1~\mathrm{MHz}$.}
    \label{FIG2}
\end{figure*}
The three-body interaction originates from the influence of magnetic flux induced by the YIG sphere on the spin-dependent energy of ASQ. Based on the phase relationship $\hat\varphi_1=\hat\varphi+\pi\Phi_\mathrm{ext}/\Phi_0$~\cite{SM}, the first term in Eq.~(\ref{EQ1}) can be expressed as
$-E_\mathrm{SO}\boldsymbol\sigma\cdot\boldsymbol n\sin{\left(\hat{\varphi}+\phi_\text{ext}\right)}$. In the subspace $\{\ket{\uparrow},\ket{\downarrow}\}$, the spin along the polarization direction reads $\boldsymbol\sigma\cdot\boldsymbol n=\cos{\theta}\hat{\sigma}^z_a+\sin{\theta}\hat{\sigma}^x_a$, where $\theta$ is the angle between the direction of the external Zeeman field and the polarization of the spin~\cite{PRXQuantum.6.010308,SM}.
Assuming the angle $\theta=\pi/2$ and considering the influence of the magnetic flux induced by the YIG sphere $\phi_\text{ext}\rightarrow \phi_\text{ext}+\phi(\hat{\boldsymbol\mu})$, the interaction term can be rewritten as $
\hat{H}_\text{int}=-E_\mathrm{SO}\hat{\sigma}^x_a\sin(\phi_\text{ext}+\hat{\varphi}+\phi(\hat{\boldsymbol\mu}))  $,
where $\phi(\hat{\boldsymbol\mu})=\pi\Phi(\hat{\boldsymbol\mu})/\Phi_0$. Expanding the trigonometric function with $\phi(\hat{\boldsymbol\mu})\ll1$ for typical parameters, the Hamiltonian becomes
\begin{equation}
    \hat{H}_{\mathrm{int}}=-E_\text{SO}{\phi(\hat{\boldsymbol\mu})}\cos({\phi_\mathrm{ext}}+\hat{\varphi})\hat{\sigma}^x_a-E_\text{SO}\sin({\phi_\mathrm{ext}} +\hat{\varphi})\hat{\sigma}^x_a.
    \label{eq3}
\end{equation}
In terms of the transmon phase operator $\hat{\tilde{\varphi}}=\hat{\varphi}-\varphi_0$, the first term in Eq.~\ref{eq3} becomes $\hat{H}_{\mathrm{int1}}=    {E_\mathrm{SO}\phi(\hat{\boldsymbol\mu})\hat{\sigma}^x_a}/{S(\phi_\text{ext})} [ {(1+a)\sin{2\phi_\mathrm{ext}}}\sin{\hat{\tilde{\varphi}}}/{2}+(\cos^2{\phi_\mathrm{ext}}-a\sin^2\phi_\mathrm{ext})\cos{\hat{\tilde{\varphi}}}]$. Expanding to order $\hat{\tilde{\varphi}}^2$ and substituting the quantized flux $\Phi(\hat{\boldsymbol{\mu}})$ and the phase operator $\hat{\tilde{\varphi}}$ with the corresponding magnon bosonic operator $\hat{m}$ $(\hat{m}^{\dagger})$ and the SCQ Pauli operator $\hat{\sigma}_s^{+}$ $(\hat{\sigma}_s^{-})$, the three-body interaction is obtained, and can be expressed as
\begin{equation}
\hat{H}_{\mathrm{thr}}/\hbar= G(\hat{\sigma}^+_s+\hat{\sigma}^-_s)(\hat{m}+\hat{m}^\dagger)\hat{\sigma}^x_a
+ J\hat{\sigma}^+_s \hat{\sigma}^-_s(\hat{m}+\hat{m}^\dagger)\hat{\sigma}^x_a.
\label{eq4}
\end{equation}
There are also two-body interactions $\hat H_\mathrm{two}$ from the second term in Eq.~\ref{eq3}~\cite{SM}. The total Hamiltonian of the hybrid quantum system is $\hat{H}_\mathrm{tot}=\hat{H}_M+\hat{H}_\mathrm{ASQ}+\hat{H}_\mathrm{SCQ}+\hat{H}_\mathrm{thr}+\hat{H}_\mathrm{two}$. Under the resonance condition $\omega_a=\omega_m+\omega_s$ (or $\omega_m=\omega_a+\omega_s$), the two-body interaction $\hat{H}_\mathrm{two}$ and the second term of the three-body interaction in Eq.~\ref{eq4} are far off-resonance and can be neglected~\cite{SM}. Applying the rotating wave approximation, the effective Hamiltonian thus is reduced to
\begin{equation}\label{Heff}
\hat{H}_\mathrm{eff}=\hat{H}_M+\hat{H}_\mathrm{ASQ}+\hat{H}_\mathrm{SCQ}+\hbar G(\hat{m}^\dagger\hat{\sigma}^+_s\hat{\sigma}^-_a+\hat{m} \hat{\sigma}^-_s\hat{\sigma}^+_a),
\end{equation}
with $\omega_a=\omega_m+\omega_s$. For the resonance condition $\omega_m=\omega_a+\omega_s$, the interaction becomes $G(\hat{m} \hat{\sigma}^+_s\hat{\sigma}^+_a+\hat{m}^\dagger \hat{\sigma}^-_s\hat{\sigma}^-_a)$.

In this work, we primarily focus on the three-body interaction, which describes the coherent exchange between the magnon, ASQ, and SCQ. For example, when the resonance condition $\omega_a=\omega_m+\omega_s$ $(\omega_m=\omega_a+\omega_s)$ is selected, the interaction $\hat{m}^\dagger \hat{\sigma}^+_s\hat{\sigma}^-_a+\hat{m} \hat{\sigma}^-_s\hat{\sigma}^+_a$ ($\hat{m} \hat{\sigma}^+_s\hat{\sigma}^+_a+\hat{m}^\dagger \hat{\sigma}^-_s\hat{\sigma}^-_a$) results in the transmission between $\ket{n,\uparrow,g}$ and $\ket{n+1,\downarrow,e}$ ($\ket{n+1,\downarrow,g}$ and $\ket{n,\uparrow,e}$). This describes magnon and SCQ excitations upon ASQ deexcitation (two qubits excitations upon magnon annihilation) and the inverse process. Its coupling strength reads
\begin{equation}
    G =\frac{E_\text{SO}\sin2\phi_\text{ext}(1+a)\pi\Phi_\text{YIG}}{2\hbar\Phi_0S(\phi_{ext})^{5/4}}(\frac{2E_C}{E_{J}^{\mathrm{sum}}})^{1/4}.
\end{equation}

 In the following, we investigate the influence of relevant parameters on the coupling strength $G$. As shown in Figs.~\ref{FIG2} (a), the coupling strength $G$ is plotted as functions of the external flux $\Phi_\text{ext}$ with different asymmetry $a$. Here, we consider spin-dependent energy $E_\text{SO}/h=600~\mathrm{MHz}$~\cite{pita2023direct,PhysRevLett.131.097001,doi:10.1126/science.abf0345}, typical transmon parameter $E_{J}^{\mathrm{sum}}/E_C=50$~\cite{RevModPhys.93.025005}, and radius of YIG sphere $R_K=30~\mu m$. The optimal point of the coupling strength $G$ depends on $a$ and its maximum value increases as $a$ decreases. $G$ varies periodically with $\Phi_\mathrm{ext}/\Phi_0$, and when $\Phi_\mathrm{ext}/\Phi_0=0.5$, $G$ is equal to zero. According to the contour maps shown in Figs.~\ref{FIG2} (b), the coupling strength $G$ is proportional to $E_\text{SO}$ and decreases with increasing $E_{J}^{\mathrm{sum}}/E_C$. However, the coupling strength $G$ is less sensitive to the ratio $E_{J}^{\mathrm{sum}}/E_C$ due to $G\propto(E_C/E_{J}^{\mathrm{sum}})^{1/4}$.

 Strong coupling is a prerequisite for conducting quantum manipulation. We adopt the decay rates of magnons, ASQs, SCQs as $\kappa_m/2\pi=0.5~\mathrm{MHz}$~\cite{shen2025cavity,PhysRevLett.132.116701}, $\kappa_a/2\pi=0.01~\mathrm{MHz}$~\cite{pita2023direct}, and $\kappa_s/2\pi=0.05~\mathrm{MHz}$~\cite{kjaergaard2020superconducting}. The dephasing rates of ASQs and SCQs are taken as $\gamma_a/2\pi=1~\mathrm{MHz}$~\cite{pita2023direct} and $\gamma_s/2\pi=0.05~\mathrm{MHz}$~\cite{kjaergaard2020superconducting}. When choosing $a=0.3$ and $\Phi_\mathrm{ext}/\Phi_0=0.35$, we have $G/2\pi\simeq 1.5~\mathrm{MHz}$. So the coupling strength $G$ can reach the strong coupling regime, i.e., $G>\kappa_m,\kappa(\gamma)_a, \kappa(\gamma)_s$. Furthermore, the coupling strength can be enhanced exponentially by the parametric amplification technique~\cite{SM}. The dephasing rate $\gamma_a$ of ASQs is expected to decrease by selecting better material for quantum dot junction to reduce the influence of the nuclear environment~\cite{pita2023direct}. The decay rate of magnons $\kappa_m=\omega_m\alpha_G$ can be reduced to $0.1~\mathrm{MHz}$ with the Gilbert-damping parameter $\alpha_G=10^{-5}$~\cite{PhysRevLett.113.083603}. Therefore, the three-body coupling strength $G$ holds promise for being much greater than the dissipations in this hybrid quantum system.

We now show that the quantum dynamics of the hybrid quantum system is governed by the three-body interaction $ G(\hat{m}^\dagger\hat{\sigma}^+_s\hat{\sigma}^-_a+\hat{m} \hat{\sigma}^-_s\hat{\sigma}^+_a)$ under the resonance condition $\omega_a=\omega_m+\omega_s$, through numerically solving the quantum master equation. To verify the validity of neglecting the far off-resonant interactions, both the effective Hamiltonian $\hat{H}_\mathrm{eff}$ and the total Hamiltonian $\hat{H}_\mathrm{tot}$ are used to simulate the population evolution of the magnon, ASQ, and SCQ with the initial state $\ket{0,\uparrow,g}$. Fig.~\ref{FIG2}(c) shows almost-perfect joint excitation of  the magnon mode and SCQ by the ASQ, i.e., if the magnon mode gets excited, the SCQ is also in the excited state. When the dissipation rates of ASQs and magnons are reduced, the populations of the SCQ and the magnon mode approach one after a half Rabi period. We also observe the excellent agreement between the dynamics generated by $\hat{H}_{\mathrm{eff}}$ and $\hat{H}_{\mathrm{tot}}$, confirming that these far off-resonant interactions can be safely neglected.
Furthermore, the population evolution under the interaction $G(\hat{m} \hat{\sigma}^+_s\hat{\sigma}^+_a+\hat{m}^\dagger \hat{\sigma}^-_s\hat{\sigma}^-_a)$ shows almost-perfect joint excitation of two qubits by magnons, i.e., if one qubit gets excited, the other one is also in the excited state~\cite{SM}.

\begin{figure}[t]
    \centering
    \includegraphics[width=0.5\textwidth]{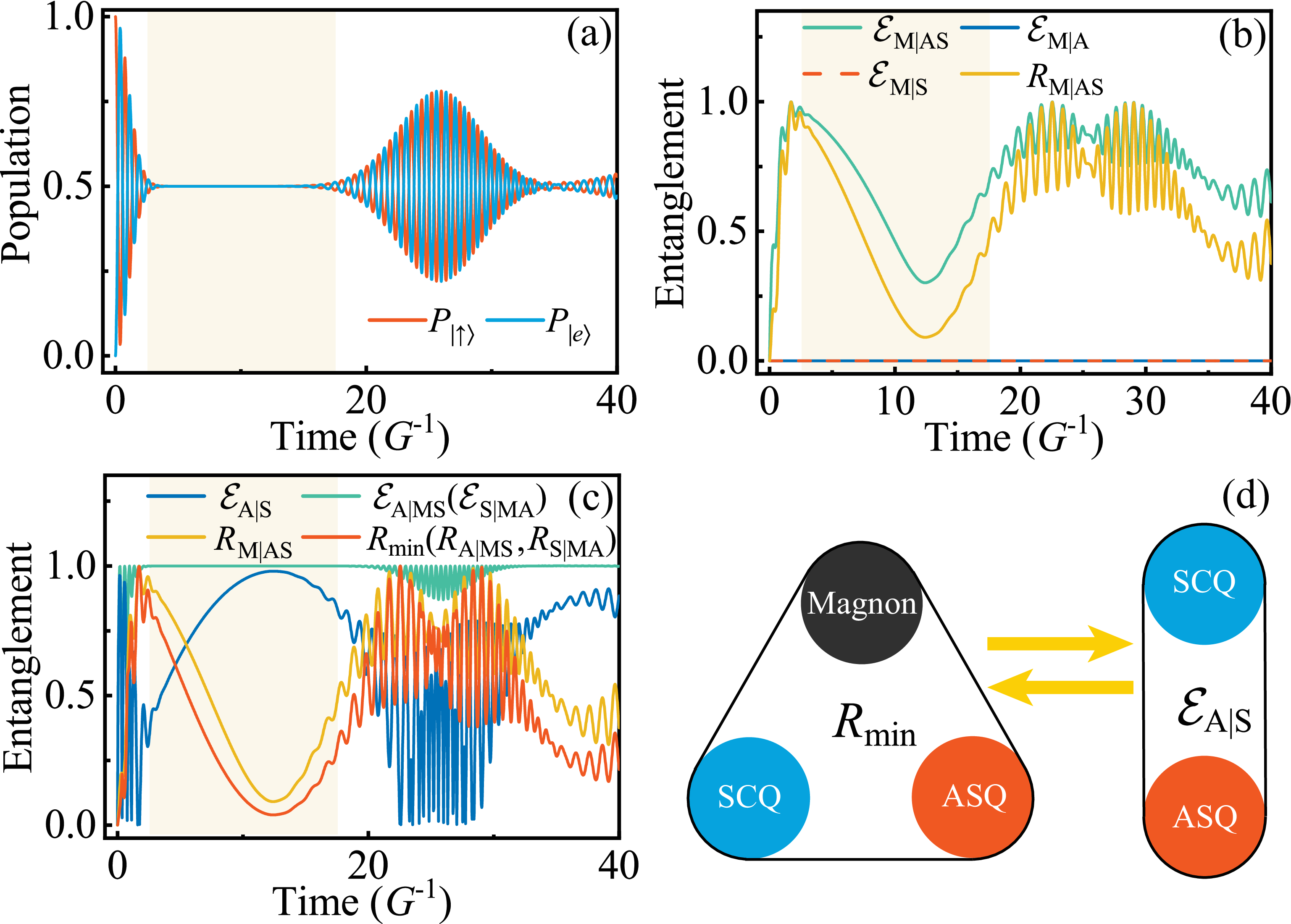}
    \caption{(a) Simulated evolution of populations without
 decoherence. The shaded region denotes the collapse region. (b) and (c) shows the evolution of the bipartite entanglement and the residual entanglements with time. (d) schematic diagram of entanglement redistribution.}
    \label{FIG3}
\end{figure}

\textit{Collapse-revival phenomena and entanglement redistribution.}---Collapse and revival~\cite{PhysRevLett.44.1323,PhysRevLett.65.3385,PhysRevA.45.8190,PhysRevLett.96.050502,PhysRevLett.96.050502,PhysRevLett.105.263603} is a well-known phenomenon based on the Jaynes-Cummings (JC) interaction, involving the disappearance of oscillations from quantum phase disorder (collapse) and their recovery via phase reordering (revival), which has been observed experimentally~\cite{doi:10.1126/science.adf7553,PhysRevLett.58.353,PhysRevLett.76.1800}. In the model of three-body interaction, the collapse and revival still exist, yet more interesting new phenomena have been observed in the dynamic process. We consider the three-body interaction $G(\hat{m}^\dagger\hat{\sigma}^+_s\hat{\sigma}^-_a+\hat{m} \hat{\sigma}^-_s\hat{\sigma}^+_a)$ when $\omega_a=\omega_m+\omega_s$. The magnon is prepared in the coherent state $\ket{\alpha}$ with $\alpha=4$, the ASQ and SCQ are prepared in the excited state $\ket{\uparrow}$ and ground state $\ket{g}$, respectively. The populations of the ASQ $\ket{\uparrow}$ state and the SCQ $\ket{e}$ state are analytically calculated to study the collapse-revival phenomena. The results are as follows: $P_{\ket{\uparrow}} = 0.5 + 0.5X,~P_{\ket{e}}= 0.5 - 0.5 X,$
where $X=\sum_{n = 0}^{\infty}C_n\cdot \cos\left( 2  \Omega_nt \right)$ with Rabi frequency $\Omega_n=G\sqrt{n+1}$ and $C_n=e^{-|\alpha|^2 } \frac{\alpha^{2n}}{{n!}}$. As shown in Fig.~\ref{FIG3}(a), when these components cancel each other out due to random phases, the series summation $X=0$, the oscillations of $ P_{\ket{\uparrow}}$ and $P_{\ket{e}}$ disappear and the populations remain at $0.5$. When these components achieve phase synchronization, their superposition reaches a maximum, manifesting as the revival of oscillations.

Beyond the collapse and revival of populations, an intriguing evolution of entanglement emerges from the dynamics of the three-body interaction. We now show that during the collapse region, where the populations remain stationary, the entanglement structure undergoes a dramatic and continuous reorganization. For the three-body quantum system, the entanglement quantified by the logarithmic negativity satisfies the CKW monogamy inequality~\cite{PhysRevA.61.052306}:
$\mathcal{E}_{i|jk}^2 \geq \mathcal{E}_{i|j}^2+\mathcal{E}_{i|k}^2$ $(i,j,k=\mathrm{M,S,A})$, where the logarithmic negativity $\mathcal{E}_{i|j}$ ($\mathcal{E}_{i|jk}$) measures the bipartite entanglement between $i$ and $j$ ($i$ and the single object $jk$). Herein, $\mathrm{M}$, $\mathrm{A}$, and $\mathrm{S}$ represent the magnon, ASQ, and SCQ, respectively. The difference between the two sides of the above inequality is the residual entanglement
\begin{equation}
    R_{i|jk}=\mathcal{E}_{i|jk}^2-\mathcal{E}_{i|j}^2-\mathcal{E}_{i|k}^2,
\end{equation}
which serves to quantify tripartite entanglement that cannot be accounted for by any bipartite divisions~\cite{PhysRevA.75.062308,PhysRevLett.113.240501,PhysRevLett.106.190502,PhysRevLett.121.203601}. First, we investigate the residual entanglement for the bipartition between the magnon (M) and the remaining two-qubits partition (AS). As shown in Fig.~\ref{FIG3}(b), $\mathcal{E}_\mathrm{M|A}$ and $\mathcal{E}_\mathrm{M|S}$ remain zero at all times, meaning that there is no bipartite entanglement between the magnon and each qubit, while such bipartite entanglement between the magnon and  qubit is present in the JC model~\cite{SM}. So the residual entanglement for the bipartition $\mathrm{M|AS}$ becomes
\begin{equation}
    R_\mathrm{M|AS}=\mathcal{E}_\mathrm{M|AS}^2,
\end{equation}
which means that the entanglement between the magnon and the rest of the system contains only residual entanglement. The variation of residual entanglement $R_\mathrm{M|AS}$ mirrors that of $\mathcal{E}_\mathrm{M|AS}$, which gradually decreases to a minimum and then increases, as depicted in Fig.~\ref{FIG3}(b).

Since $\mathcal{E}_\mathrm{M|A}~(\mathcal{E}_\mathrm{M|S}=0)$, we can also obtain the relation for the bipartition $\mathrm{A|MS}~(\mathrm{S|MA})$ as
\begin{equation}
R_\mathrm{A|MS}~(R_\mathrm{S|MA})+\mathcal{E}_\mathrm{A|S}^2=\mathcal{E}_\mathrm{A|MS}^2~(\mathcal{E}_\mathrm{S|MA}^2),\\
\label{eq8}
\end{equation}
which indicates that the entanglement between any qubit and the rest of the system includes the bipartite entanglement between two qubits as well as the residual entanglement. Surprisingly, the entanglement between any qubit and the rest of the system exhibits collapse and revival. As shown in Fig.~\ref{FIG3}(c), in the collapse region, $\mathcal{E}_\mathrm{A|MS}~(\mathcal{E}_\mathrm{S|MA})$ is conserved and equal to 1, meaning the entanglement remains maximized. So Eq.~\ref{eq8} becomes $\mathcal{E}_\mathrm{A|S}^2+R_\mathrm{A|MS}~(R_\mathrm{S|MA})=1$ in the collapse region. The conservation of total entanglement forces a dynamic balance: the growth of residual entanglement must be compensated by the consumption of bipartite entanglement between two qubits. As shown in Fig.~\ref{FIG3}(c), the residual entanglement $R_\mathrm{A|MS}~(R_\mathrm{S|MA})$ decreases as $\mathcal{E}_\mathrm{A|C}$ increases and vice versa.
In particular, $R_\mathrm{A|MS}~(R_\mathrm{S|MA})$ coincides with the minimum residual entanglement $R_\mathrm{min}=\min[R_\mathrm{M|AS},R_\mathrm{A|MS},R_\mathrm{S|MA}]$~\cite{Adesso_2007,Adesso_2006}, since $R_\mathrm{A|MS}(R_\mathrm{S|MA})\leq R_\mathrm{M|AS}$ always holds in Fig.~\ref{FIG3}(c). The nonzero minimum residual entanglement $R_\mathrm{min}$ signifies the existence of genuine tripartite entanglement within the system. Therefore, a conclusion can be drawn regarding the entanglement dynamics in the collapse region: the total entanglement between any qubit and the rest of the system is conserved, yet its form is dynamically transformed from genuine tripartite entanglement into bipartite entanglement between the two qubits and back, as illustrated in Fig.~\ref{FIG3}(d).

This finding provides unequivocal evidence that energy and entanglement dynamics can become independent. In the collapse region, the energy remains constant, while the entanglement is redistributed from the tripartite form to the bipartite form and vice versa. This highlights the potential of three-body interaction for exploring complex entanglement dynamics beyond the reach of simple JC interaction, where such entanglement redistribution does not exist~\cite{SM}. We have also analytically calculated the bipartite entanglement measured by von Neumann entropy or concurrence, and demonstrated that the redistribution of entanglement still exists even in the presence of dissipation in the Supplemental Material~\cite{SM}.

\textit{Conclusion.}---We propose a scheme, leveraging the intrinsic spin-supercurrent coupling of the ASQ, to achieve strong three-body coupling in a system consisting of a YIG sphere, an Andreev spin qubit, and a superconducting qubit via the magnetic flux from the YIG sphere. The three-body interaction can realize the joint excitation of two qubits via single magnons, that is, two qubits can be manipulated simultaneously by a magnon. Moreover, by preparing the magnon in a coherent state, we investigate the collapse and revival of the two qubit populations. Notably, in the collapse region where the populations remain stationary, we find continuous entanglement redistribution from tripartite to bipartite forms and back with the total entanglement conserved. Our work provides a new platform for realizing strong three-body interactions, enabling the exploration of novel quantum phenomena, which is a crucial step towards multipartite hybrid architectures.
\begin{acknowledgments}
This work is supported by the National Natural Science Foundation of China under Grants No. W2411002 and No. 12375018.
\end{acknowledgments}

\end{bibunit}

\clearpage
\pagestyle{empty}
\onecolumngrid
\vspace*{10pt}
\renewcommand{\theequation}{S\arabic{equation}}
\setcounter{equation}{0}
\begin{bibunit}[apsrev4-2]
\begin{center}
	\large \textbf{Supplemental Material for ``Three-body interaction in a magnon-Andreev-superconducting qubit system: collapse-revival phenomena and entanglement redistribution"}
\end{center}

In this Supplemental Material, we provide detailed derivations and extended analyses to support the main text. Sec.~\ref{secI} introduces the Hamiltonians and quantization of the hybrid system components: the Andreev spin qubit, superconducting qubit, and magnon mode. Sec.~\ref{secII} provides the complete derivation of both three-body and two-body interactions mediated by the magnon-induced magnetic flux, demonstrating that two-body terms become negligible under specific resonance conditions. Sec.~\ref{secIII} introduces parametric amplification techniques to exponentially enhance the three-body coupling strength through microwave driving and the magnon Kerr effect. Sec.~\ref{secIV} provides an in-depth analysis of collapse-revival phenomena and entanglement redistribution, including analytical solutions for population dynamics, entanglement measures, effects of system dissipation, and comparison with Jaynes-Cummings models.

\section{\label{secI}The hybrid Magnon-Qubits system}

    In this section, we introduce the components in this hybrid quantum system, including the Andreev spin qubit (ASQ), the superconducting qubit (SCQ), and the magnon. First, we show the realization of ASQ in the quantum dot junction. Then, we calculate the Hamiltonian of SCQ through the inductive energy of the SUQID and the energy of the capacitor. Finally, we demonstrate the quantization of the spin wave in a ferromagnetic microsphere and obtain the Hamiltonian of the magnon modes.

    \subsection{ The Andreev spin qubit}
\begin{figure}[tp]
        \centering
    \includegraphics[width=0.6\columnwidth]{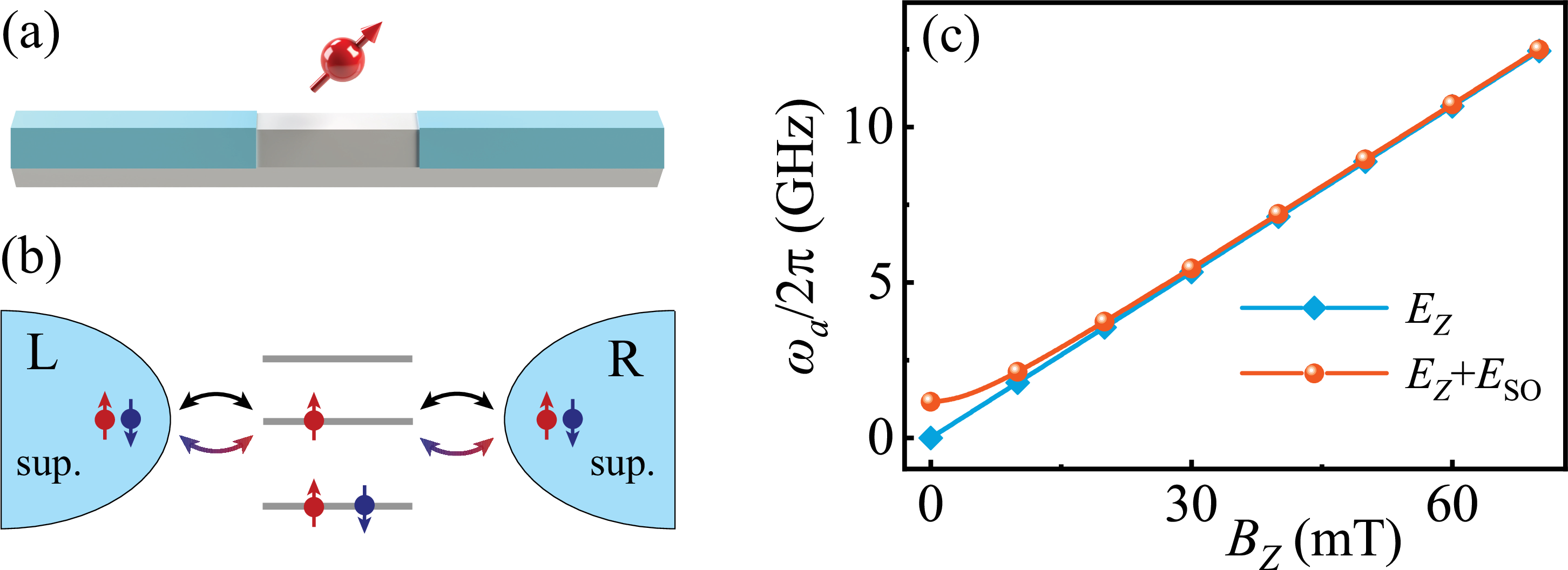}
    \caption{\label{figflux}(a)  Illustration of a semiconductor (silver) with epitaxial
     superconducting leads (light blue). (b) Conceptual diagram of cooper pair tunneling between semiconducting quantum dot and superconductors. Black and gradient color denote the spin-conserving and spin-flipping tunnel, respectively. (c) Plot of the frequency of the ASQ as functions of the external magnetic field $B_Z$. The angle and phase difference are $\theta=\pi/2$ and $\varphi_1=\pi/2$.}
    \label{f1}
    \end{figure}

    The Andreev spin qubit is implemented using semiconducting quantum dot Josephson junction. As shown in Fig.~\ref{f1}(a), the semiconducting quantum dot Josephson junction is composed of a hybrid semiconducting-superconducting nanowire, where the semiconductor and the superconducting leads are usually made of InAs and Al in the experiment~\cite{doi:10.1126/science.abf0345,pita2023direct,pita2024strong}. By adjusting the gate voltages to control the occupation of the quantum dot, the ground-state manifold of the quantum dot junction is composed of two spin states. The energy of the quantum dot junction originates from Cooper pair tunneling via a series of single-electron cotunneling processes through the energy levels of the quantum dot, as illustrated in Fig.~\ref{f1}(b). Spin-conserving tunneling processes, in which both electrons cotunnel through the same energy level, result in the spin-independent Josephson energy $E_0$. Spin-flipping tunneling processes, in which one electron cotunnels through the singly occupied level (involving a spin rotation) while the second one cotunnels through a different level, result in the spin-dependent energy $E_\mathrm{SO}$. Here, the single-electron tunneling amplitudes are spin-dependent due to the presence of spin-orbit coupling. In the presence of the external magnetic field $\boldsymbol{B}_Z$ applied to the quantum dot junction, Zeeman energy can be achieved via the Zeeman effect, and it can be expressed as
    \begin{equation}
     U_Z=\frac{1}{2}\boldsymbol{E}_Z\cdot\boldsymbol{\sigma}
    \end{equation}
    where $\boldsymbol{E}_Z=\hbar g_L\mu_B\boldsymbol{B}_Z$ is the Zeeman field with Land\'{e} factor $g_L=12.7$ and Bohr magneton $\mu_B$.
    The total potential energy of the quantum dot junction is then expressed as
    \begin{equation}
     U_\mathrm{tot}=-E_0\cos{\varphi_1}-E_\mathrm{SO}\boldsymbol{\sigma}\cdot\boldsymbol{n}\sin{\varphi_1}+\frac{1}{2}\boldsymbol{E}_Z\cdot\boldsymbol{\sigma}
    \end{equation}
    where $\varphi_1$ is the phase difference across the quantum dot junction, $\boldsymbol{\sigma}$ is the spin operator, $\boldsymbol{n}$ is the unit vector along spin-polarization direction. The realization of the ASQ hinges on the spin-dependent energy $E_\mathrm{SO}$ and Zeeman energy, and its Hamiltonian is expressed as
    \begin{equation}
     H_\mathrm{ASQ}=-E_\mathrm{SO}\boldsymbol{\sigma}\cdot\boldsymbol{n}\sin{\varphi_1}+\frac{1}{2}\boldsymbol{E}_Z\cdot\boldsymbol{\sigma}
    \end{equation}
    We define the direction of the Zeeman field as the z-direction of the spin. The Hamiltonian of ASQ can be written as
    \begin{align}
    \hat{H}_\mathrm{ASQ}&=-E_\mathrm{SO}\sin{\varphi_1}(\cos{\theta}\hat{\tilde{\sigma}}^z_a+\sin
     {\theta}\hat{\tilde{\sigma}}^x_a)+\frac{1}{2}{E}_Z\hat{\tilde{\sigma}}_a^z\nonumber\\
     &=\frac{1}{2}\begin{pmatrix}
     E_Z-2E_\mathrm{SO}\sin{\varphi_1}\cos{\theta} & -2E_\mathrm{SO}\sin{\varphi_1}\sin{\theta} \\
    -2E_\mathrm{SO}\sin{\varphi_1}\sin{\theta} & -{E}_Z+2E_\mathrm{SO}\sin{\varphi_1}\cos{\theta}
     \end{pmatrix}
    \end{align}
    where $\theta$ denotes the angle between the direction of the external
    Zeeman field and the spin-polarization of the spin, $\hat{\tilde{\sigma}}^z_a=\tilde{\ket{\uparrow}}\tilde{\bra{\uparrow}}-\tilde{\ket{\downarrow }}\tilde{\bra{\downarrow }}$ and $\hat{\tilde{\sigma}}_a^x=\tilde{\ket{\uparrow}}\tilde{\bra{\downarrow}}+\tilde{\ket{\uparrow }}\tilde{\bra{\downarrow }}$ denote the Pauli z and Pauli x operator in the subspace $\{\tilde{\ket{\uparrow}},\tilde{\ket{\downarrow}}\}$, respectively. Diagonalizing the Hamiltonian, the eigenvalues can be expressed as
    \begin{equation}
        \lambda_{\pm} = \pm \frac{1}{2} \sqrt{E_Z^2 - 4 E_Z E_{\mathrm{SO}} \sin\varphi_1 \cos\theta + 4 E_{\mathrm{SO}}^2 \sin^2\varphi_1}.
    \end{equation} In the limit of $E_Z\gg E_\mathrm{SO}$, we obtain $\lambda_{\pm}\simeq\pm \frac{1}{2}E_Z $, and the eigenstates of the ASQ as $\ket{\uparrow} \simeq \tilde{\ket{{\uparrow}}}$ and $\ket{\downarrow} \simeq \tilde{\ket{{\downarrow}}}$. The Hamiltonian of ASQ simplifies to
    \begin{equation}
        \hat H_\mathrm{ASQ}=\frac{\hbar}{2}\omega_a\hat{\sigma}_a^z,
    \end{equation}
    where $\omega_a=E_Z/\hbar$ is the frequency of ASQ, and $\hat{\sigma}_a^z={\ket{\uparrow}}{\bra{\uparrow}}-{\ket{\downarrow }}{\bra{\downarrow }}$ is the Pauli z operator in the new $\{\ket{\downarrow },\bra{\downarrow }\}$ basis. Similarly, in the new subspace $\{\ket{\downarrow },\bra{\downarrow }\}$, $\boldsymbol{\sigma}\cdot\boldsymbol{n}=(\cos{\theta}\hat\sigma^z_a+\sin{\theta}\hat\sigma^x_a)$ remains unchanged. As shown in Fig.~\ref{f1}(c), we plot the variation of frequency $\omega_a$ with magnetic field $B_Z$ for the case of considering only the influence of $E_Z$ and considering the combined influence of $E_Z$ and $E_{\mathrm{SO}}$. It can be seen that for large magnetic fields $B_Z$, the contribution of the terms with $E_{\mathrm{SO}}$ to the frequency can be negligible.
    \subsection{ The superconducting qubit}
\begin{figure}[tp]
        \centering
    \includegraphics[width=0.6\columnwidth]{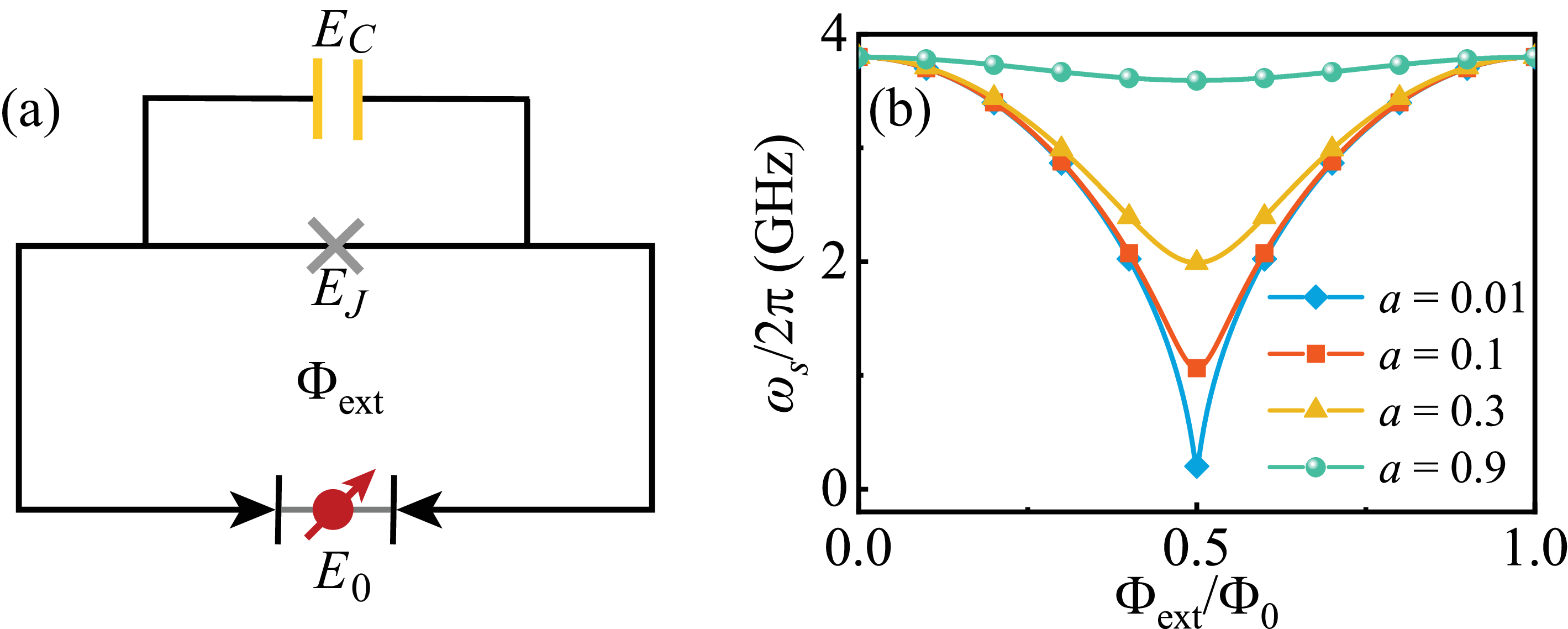}
    \caption{\label{figflux}(a) Superconducting circuit model of capacitor, and the SQUID consisting of the quantum dot junction and the conventional Josephson. (b) The frequency $\omega_S$ of the SCQ versus the external flux $\Phi_\mathrm{ext}$ for different junction asymmetry $a$. The relevant parameters are selected as $E_C/h=200\mathrm{MHz}$ and $E_J^\mathrm{sum}/h=10\mathrm{GHz}$}
    \label{f2}
    \end{figure}

    A superconducting quantum interference device (SQUID), as depicted in Fig.~\ref{f2}, is constructed by connecting a conventional Josephson junction (Josephson energy $E_J$) and a quantum dot junction (Josephson energy $E_0$) in parallel, as shown in Fig.~\ref{f2}. The inductive energy of SQUID reads $\hat H_\mathrm{ind}=-E_{0}\cos{( \hat{\varphi}_1)}-E_{J}\cos{( \hat{\varphi}_2)}$, where $\hat\varphi_1$ and $\hat\varphi_2$ represent the phase difference across the  quantum dot junction and the  conventional Josephson junction. When an external magnetic flux $\Phi_\mathrm{ext}$ passes through the loop, flux quantization requires that
    \begin{equation}
     \hat\varphi_1-\hat\varphi_2=2\pi\Phi_\text{ext}/\Phi_0.
    \end{equation}
    Define the average phase difference as $\hat{\varphi}=(\hat{\varphi}_2+\hat{\varphi}_1)/2$, the phase difference $\hat\varphi_1$ and $\hat\varphi_2$ can be expressed
    \begin{equation}
    \hat\varphi_1=\hat\varphi+\pi\Phi_\text{ext}/\Phi_0,\quad\hat\varphi_2=\hat\varphi-\pi\Phi_\text{ext}/\Phi_0.
    \label{es8}
    \end{equation}
    The inductive energy of SQUID can then be rewritten as
    \begin{equation}
    \hat H_\mathrm{ind}=-E_{J}^\mathrm{sum}S(\phi_\mathrm{ext})\cos{( \hat{\varphi}-\varphi_0 )},
    \end{equation}
    where $E_{J}^\mathrm{sum}=E_J+E_{0}$, ${S(\phi_\mathrm{ext})=\sqrt{\cos^{2}{\phi_\mathrm{ext}}+a^{2}\sin^{2}{\phi_\mathrm{ext}}}}$ with junction asymmetry $a=(E_J-E_0)/E_{J}^{\mathrm{sum}}$ and phase $\phi_\mathrm{ext}=\pi\Phi_\mathrm{ext}/\Phi_0$, and $\varphi_0=\arctan(a\tan{\phi_\mathrm{ext}})$. We can obtain a superconducting qubit (SCQ) by the SQUID loop in parallel to a capacitor whose energy is $4E_{C}\hat{N}^{2}$ with the charging energy $E_C$ and the charge number operator $\hat{N}$~\cite{doi:10.1126/science.239.4843.992,martinis2020quantum,RevModPhys.93.025005}. The Hamiltonian of the superconducting qubit can be written as
    \begin{equation}
       \hat{H}_\mathrm{SCQ}=4E_{C}\hat{N}^{2} -E_{J}^\mathrm{sum}S(\phi_\mathrm{ext})\cos{( \hat{\varphi}-\varphi_0 )}
    \end{equation}
    In terms of the bosonic operators $\hat{s}$ ($\hat{s}^\dagger$), $\hat{N}$ and $\hat{\tilde{\varphi}}$  can be expressed
    \begin{equation}
     \hat{N}=\frac{i}{2}(\frac{E_{J}^\mathrm{sum}S(\phi_\mathrm{ext})}{2E_C})^{1/4}(\hat{s}^\dagger-\hat{s}),\quad \hat{\tilde{\varphi}}=(\frac{2E_C}{E_{J}^\mathrm{sum}S(\phi_\mathrm{ext})})^{1/4}(\hat{s}^\dagger+\hat{s}),
     \label{eq11}
    \end{equation}
    where $\hat{\tilde{\varphi}}=\hat{\varphi}-\varphi_0 $ is the transmon phase operator. In the transmon regime ($E_{J}^\mathrm{sum}S(\phi_\mathrm{ext})\gg E_C$)~\cite{RevModPhys.73.357,PhysRevA.76.042319,clarke2008superconducting}, the zero-point fluctuation $\hat{\tilde{\varphi}}_\mathrm{zpf}=({2E_C}/{E_{J}^\mathrm{sum}S(\phi_\mathrm{ext})})^{1/4}$ of the phase variable is very small, which leads to the approximate Hamiltonian of SCQ
    \begin{align}
       \hat{H}_\mathrm{SCQ}&=4E_{C}\hat{N}^{2} -E_{J}^\mathrm{sum}S(\phi_\mathrm{ext})(\frac{\hat{\tilde{\varphi}}^2}{2}-\frac{\hat{\tilde{\varphi}}^4}{24}) \nonumber\\
       &=\hbar\omega_s \hat{s}^{\dagger}\hat{s} - \frac{E_C}{2} \hat{s}^{\dagger}\hat{s}^{\dagger}\hat{s}\hat{s}.
    \end{align}
    Here, $\omega_s=(\sqrt{8 E_C E_J^\mathrm{sum} S(\phi_\mathrm{ext})} - E_C)/\hbar$ is the frequency of SCQ. The variations of frequency $\omega_s$ with external flux $\Phi_\mathrm{ext}$ for different junction asymmetry $a$ are plotted in Fig.~\ref{f2}(b). It is noteworthy that because of the system's significant anharmonicity $-E_C$, the bosonic operators $\hat{s}$ and $\hat{s}^\dagger$ can be mapped to the Pauli lowering and raising operators under the two-level approximation, i.e.,
    \begin{equation}
        \hat{s}\rightarrow\hat{\sigma}^-_s,\quad\hat{s}^\dagger\rightarrow\hat{\sigma}^+_s.
    \end{equation}
    As a result, the Hamiltonian reduces to an effective qubit form:
    \begin{equation}
        \hat{H}_\mathrm{SCQ}=\frac{\hbar\omega_s}{2} \hat{\sigma}_s^z,
    \end{equation}
    where $\hat{\sigma}_s^z=\ket{e}\bra{e}-\ket{g}\bra{g }$ is the Pauli z operator in the two-level subspace $\{\ket{g},\ket{e}\}$ of the ASQ.
    \subsection{The quantization of the spin wave}
    Spin waves, first theorized by Bloch, represent collective excitations in magnetically ordered systems where exchange interactions maintain parallel spin alignment~\cite{RevModPhys.30.1}. When a bias magnetic field is applied to magnetic systems, the spins will be disturbed and deviate from the parallel state, resulting in a deflection angle between adjacent spins. Due to dipole interaction or exchange interaction, this disturbance will be transmitted to the nearest neighbors. As the disturbance propagates, spins at different positions will all have small angle deflections and show periodic changes. Eventually, a collective excitation similar to lattice vibration will be generated, which is the spin wave.

    To describe spin waves in a ferromagnetic microsphere, we first present its Hamiltonian:

    \begin{equation}
    \hat {\mathcal{H} }= -\mathit{g}\mu_B B_K \sum_{i} \hat{S}_i^z - 2\mathcal{J} \sum_{\langle i,j \rangle} \hat{\vec{S}}_i \cdot \hat{\vec{S}}_j
    \label{eq15}
    \end{equation}
    The first term represents the Zeeman energy, and the second term represents the sum of nearest-neighbor exchange interactions. Here, $\mathit{g}$ is the $\mathit{g}$-factor, $\mu_B $ is the Bohr magneton, \( B_K \) is the static magnetic field applied to the ferromagnetic microsphere along the \( z \) axis, \( \hat{\vec{S}}_i \) is the Heisenberg spin operator at the \( i \)-th site, and $\mathcal{J}$ is the exchange integral. For ferromagnetic materials, $\mathcal{J}$ takes a positive value, leading to the ferromagnetic ground state. In the ground state, all spins are oriented along the \( z \)-axis.

    Through the Holstein-Primakoff formula~\cite{PhysRev.58.1098}, the Heisenberg spin operator \( \hat{\vec{S}}_i \) in Eq.~\ref{eq15} can be expressed in terms of bosonic operators \( \hat{c}_i \) and \( \hat{c}_i^\dagger \), as follows:

    \begin{align}
    &\hat{S}_i^+ = \hat{S}_i^x + i\hat{S}_i^y = \sqrt{2S} \left( 1 - \frac{\hat{c}_i^\dagger \hat{c}_i}{2S} \right)^{\frac{1}{2}} \hat{c}_i^\dagger,\\
    &\hat{S}_i^- = \hat{S}_i^x - i\hat{S}_i^y = \sqrt{2S} \left( 1 - \frac{\hat{c}_i^\dagger \hat{c}_i}{2S} \right)^{\frac{1}{2}} \hat{c}_i,\\
    &\hat{S}_i^z = S - \hat{c}_i^\dagger \hat{c}_i,
    \label{eq18}
    \end{align}
    where \( S \) is the spin at each spatial position. From Eq.~\ref{eq18}, it can be found that the number of magnons \( c_i^\dagger c_i \) corresponds to the decrease in the \( z \)-component \( S_i^z \) of the total spin. Through Fourier transformation, the bosonic operators \( c_i \) and \( c_i^\dagger \) can be defined by the spin wave operators at each lattice site:
    \begin{align}
    &\hat{c}_i = \frac{1}{\sqrt{N}} \sum_{k} e^{-ik \cdot r_i} \hat{c}_k\\
    &\hat{c}_i^\dagger = \frac{1}{\sqrt{N}} \sum_{k} e^{-ik \cdot r_i} \hat{c}_k^\dagger
    \end{align}

    Here, \( \hat{c}_k \) and \( \hat{c}_k^\dagger \) are the annihilation and creation operators of magnon in a certain plane wave mode, and \( N \) is the number of atoms with spin \( S \). Substitute these operators into Eq.~\ref{eq15} and truncate the terms higher than the second order. Then, the Hamiltonian after the spin waves are quantized can be written as
    \begin{equation}
    \hat {\mathcal{H} } = \sum_{k} \hbar\omega_{k} \hat{c}_{k}^{\dagger} \hat{c}_{k} .
    \end{equation}
    The corresponding dispersion relation is
    \begin{equation}
    \hbar\omega_{k} = 2SZ(1 - \gamma_{k}) + g\mu_{B}B_{K}.
    \end{equation}
    Here, \( Z \) is the coordination number of the crystal lattice sites of the corresponding material. Assuming a simple cubic structure, then \( Z = 6 \). Let the lattice constant be denoted as \( a_{0} \), then \( \gamma_{k} \) is expressed as
    \begin{equation}
     \gamma_{k} = \frac{1}{3}( \cos k_{x}a_{0} + \cos k_{y}a_{0} + \cos k_{z}a_{0} ).
    \end{equation}
    It can be derived that the quadratic dispersion relation in the long-wavelength limit is
    \begin{equation}
    \hbar\omega_{k} = 2Sa_{0}^{2}k^{2} + g\mu_{B}B_{K}.
    \end{equation}
    Observing the first term, the rigidity of the ordered spin system enhances the degeneracy of spin excitations, which forms a significant contrast to the situation of the paramagnetic spin system.

    As for the Kittel mode, all spins are in phase ($k=0$). The corresponding dispersion relation becomes
     \begin{equation}
         \hbar\omega_{m} = g\mu_{B}B_{K},
     \end{equation}
    where $\omega_m=\gamma_0B_K $ is the frequency of the Kittel mode depending on the external static field with the gyromagnetic ratio $\gamma_0$.
    So, the Hamiltonian of Kittel modes can be written as
    \begin{equation}
        \hat{H}_\mathrm{M}=\hbar\omega_m\hat{m}^{\dagger}\hat{m}.
    \end{equation}
    Here, we employ the annihilation (creation) operator of Kittel modes $\hat{m}$ $(\hat{m}^{\dagger})$.
    \section{\label{secII}Derivation of the interaction in Hybrid Magnon-Qubits Systems}
    In this section, we show the two-body and three-body interactions in this hybrid system. Frist, the magnetic flux induced by the YIG sphere is calculated. Then, we consider the influence of the magnetic flux on the original Hamiltonian and obtain the three-body interaction and all two-body interactions. Finally, we demonstrate that the coupling strengths of these off-resonant two-body interactions are much smaller than the detuning, which leads to their being negligible.
    \subsection{Magnetic flux induced by YIG sphere}
\begin{figure}[tp]
        \centering
    \includegraphics[width=0.9\columnwidth]{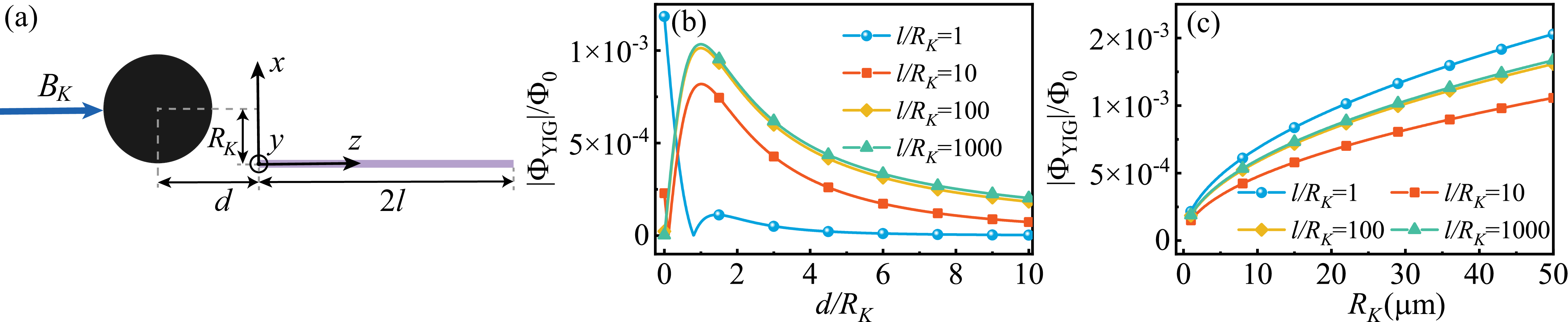}
    \caption{\label{fg3}(a) Schematic illustration of the YIG sphere and the square SQUID loop. (b) The ratio $|\Phi_\mathrm{YIG}|/\Phi_0$ versus the funvtion $d/R_K$ for different relative size $l/R_K$. (c) The ratio $|\Phi_\mathrm{YIG}|/\Phi_0$ versus the radius $R_K$ for different relative size $l/R_K$.}
    \end{figure}

    We set up a coordinate system with its origin at the midpoint of the left arm of the square SQUID loop, such that the loop lies in the $yz$ plane in Fig.~\ref{fg3}(a). Thus, any point on the loop has coordinates $(0,y,z)$. The center of the YIG sphere lies on $xz$ plane, whose coordinates are $(R_K,0,-d)$. The center of the YIG sphere is located at a distance $R_K$ in the x-direction from the origin, where $R_K$ is also the radius of the sphere. Then the position vector from the center of the YIG sphere to any location on the SQUID loop is $\boldsymbol{r}=(-R_K,y,z+d)$. In classical electrodynamics, the magnetic field generated by a magnetic sphere at position r can be described by the magnetic dipole model~\cite{PhysRevLett.124.163604}, which is given by
     \begin{align}
     \boldsymbol{B}({\boldsymbol{\mu}})=\frac{\mu_0}{4\pi r^3}[\frac{3\boldsymbol{r}(\boldsymbol{\mu}\cdot\boldsymbol{r})}{r^2}-\boldsymbol{\mu}],
     \end{align}
    where $\mu_0$ is the permeability of the free space and $\boldsymbol{\mu}$ is the magnetic moment. The magnetic moment $\boldsymbol{\mu}$ can be quantized within the Holstein-Primakoff approximation as
      \begin{align}
    \hat{\mu}_x=\hbar\gamma_0\sqrt{N_S/2}(\hat{m}^\dagger+\hat{m}),         \quad\hat{\mu}_y=i\hbar\gamma_0\sqrt{N_S/2}(\hat{m}-\hat{m}^\dagger),\quad\hat{\mu}_z=\hbar\gamma_0\hat{m}^\dagger\hat{m},
     \end{align}
    where $N_S$ is the total number of spins~\cite{PhysRevA.100.022343}. Within the Holstein-Primakoff approximation, we overlook a very small contribution from $\hat{\mu}_z$. Then the quantized magnetic field components can be expressed as
     \begin{align}
     &\hat{B}_x=\frac{\hbar\mu_0\gamma_0\sqrt{N_S/2}}{4\pi r^3}[\frac{3R_K^2(\hat{m}+\hat{m}^\dagger)-3iR_Ky(\hat{m}-\hat{m}^\dagger)}{r^2}-(\hat{m}+\hat{m}^\dagger)]\\
     &\hat{B}_y=\frac{\hbar\mu_0\gamma_0\sqrt{N_S/2}}{4\pi r^3}[\frac{-3R_Ky(\hat{m}+\hat{m}^\dagger)+3i y^2(\hat{m}-\hat{m}^\dagger)}{r^2}-i(\hat{m}+\hat{m}^\dagger)]\\
    &\hat{B}_z=\frac{\hbar\mu_0\gamma_0\sqrt{N_S/2}}{4\pi r^3}[\frac{-3R_K(z+d)(\hat{m}+\hat{m}^\dagger)+3iy(z+d)(\hat{m}-\hat{m}^\dagger)}{r^2}]
      \end{align}

     The magnetic field generated by the magnetic sphere induces a magnetic flux through the SQUID loop area ${S}$. The magnetic flux is given by the surface integral:
     \begin{align}
    \Phi(\hat{\boldsymbol{\mu}})=\iint\boldsymbol{B}(\hat{\boldsymbol{\mu}})\cdot d\boldsymbol{S}\label{flux 1}
     \end{align}
    Since the SQUID loop is located in the $yz$ plane, $\hat{B}_y\cdot d\boldsymbol{S}$ and $\hat{B}_z\cdot d\boldsymbol{S}$ are equal to zero. Eq.~\ref{flux 1} can be simplified as
    \begin{align}
     \Phi(\hat{\boldsymbol{\mu}})=\iint\hat{B}_x(\hat{\boldsymbol{\mu}})\cdot d\boldsymbol{S}=\frac{\hbar\mu_0\gamma_0\sqrt{N_S/2}}{4\pi }\int_{-l}^{l} \int_{0}^{2l}\frac{1}{r^3}[\frac{3R_K^2(\hat{m}+\hat{m}^\dagger)-3iR_Ky(\hat{m}-\hat{m}^\dagger)}{r^2}-(\hat{m}+\hat{m}^\dagger)]dy dz.\label{flux 2}
    \end{align}
    Here, $l$ is the half side length of SQUID loop. The imaginary part of the integral vanishes due to the symmetry of the SQUID loop about the $xz$ plane, which makes the integrand an odd function of y The flux can be further expressed as
    \begin{align}
     \Phi(\hat{\boldsymbol{\mu}})=\frac{\hbar\mu_0\gamma_0\sqrt{N_S/2}}{4\pi }\int_{-l}^{l} \int_{0}^{2l}\frac{1}{r^3}(\frac{3R_K^2}{r^2}-1)(\hat{m}+\hat{m}^\dagger)dy dz.\label{flux 2}
    \end{align}
    Substituting $z$ with $\tilde{z}-d$ and defining the integral of the flux
    \begin{align}
     \Phi_\mathrm{YIG}=\frac{\hbar\mu_0\gamma_0\sqrt{N_S/2}}{4\pi }\int_{-l}^{l} \int_{d}^{d+2l}\left[\frac{3R^2}{(R^2+y^2+\tilde{z}^2)^{5/2}}-\frac{1}{(R^2+y^2+\tilde{z}^2)^{3/2}} \right]dy d\tilde{z},\label{flux 2}
    \end{align}
    the flux generated by magnetic sphere is written as $\Phi(\hat{\boldsymbol{\mu}})=\Phi_\mathrm{YIG}(\hat{m}+\hat{m}^\dagger)$. As shown in Fig.~\ref{fg3}(b), the ratio $|\Phi_\mathrm{YIG}|/\Phi_0$ as a function of $d/R_K$ is plotted for different $l/R_K$. Here, $\Phi_0=h/2e$ is the magnetic flux quantum, $e$ is the absolute value of the electron charge, and h is the Planck constant. For large loop ($l\gg R_K$), $|\Phi_\mathrm{YIG}|$ get maximum in $d=R_K$, and $|\Phi_\mathrm{YIG}|$ get maximum in $d=0$ when $l \simeq R_K$. The total number of spins can be expressed as $N_S=\rho_s\frac{4\pi R_K^3}{3}$, where spin density is chosen as $\rho_s=2.14\times10^{28}/m^3$~\cite{doi:10.1126/science.aaa3693}. Selecting the optimal $d$ for different $l$, we can find $|\Phi_\mathrm{YIG}|$ increases with the radius of magnetic sphere $R_K$ in Fig.~\ref{fg3}(c).

    \subsection{The influence of the flux}
    As discussed in Sec.~\ref{secI}, the free Hamiltonian of ASQ has been derived, where we consider only a single ASQ. As the quantum dot junction and the conventional Josephson junction form a SQUID loop, under the constraint of the phase relationship in Eq.~\ref{es8}, the Hamiltonian of ASQ is rewritten as
    \begin{equation}
         \hat{{H'}}_\mathrm{ASQ}=-E_\mathrm{SO}\sin{(\hat\varphi+\pi\Phi_\mathrm{ext}/\Phi_0)}\hat{{\sigma}}_a^x+\frac{1}{2}{E}_Z\hat{{\sigma}}_a^z,
    \end{equation}
    where we assume the angle $\theta=\pi/2$. Considering the influence of the magnetic flux induced by the YIG sphere, the total magnetic flux changes to
    \begin{equation}
        \Phi_\mathrm{ext}\rightarrow \Phi_\mathrm{ext}+\Phi(\hat{\boldsymbol\mu}).
    \end{equation}
    The flux-related Hamiltonians of both ASQ and SCQ are affected, and the Hamiltonians can be reconstructed as
    \begin{align}
        \hat{\tilde{H}}_\mathrm{ASQ}&=-E_\mathrm{SO}\sin{[\phi_\mathrm{ext}+\phi(\hat{\boldsymbol\mu})+\hat\varphi]}\hat{{\sigma}}_a^x+\frac{1}{2}{E}_Z\hat{{\sigma}}_a^z,\label{eq13}\\
           \hat{\tilde{H}}_\mathrm{SCQ}&=4E_{C}\hat{N}^{2} -E_{J}^\mathrm{sum}S(\phi_\mathrm{ext}+\phi(\hat{\boldsymbol\mu}))\cos{\left\{ \hat{\varphi}-\arctan\left[a\tan({\phi_\mathrm{ext}}+\phi(\hat{\boldsymbol\mu}))\right] \right\}}\label{eq14},
    \end{align}
    where $\phi(\hat{\boldsymbol\mu})=\pi \Phi(\hat{\boldsymbol\mu})/\Phi_0$.
    \subsection{Three-body interaction and two-body interaction between magnon and ASQ}
    The three-body coupling originates from the first term in Eq.~\ref{eq13}. Expanding $\sin[\phi_{\mathrm{ext}}  - \hat{\varphi}+ \phi(\hat{\mu})]$ by the Taylor series with $\phi(\hat{\mu}) \sim10^{-3}\ll1 $, Eq.~\ref{eq13} becomes
    \begin{align}
      \hat{\tilde{H}}_\mathrm{ASQ}&=-E_\mathrm{SO}[\sin({\phi_\mathrm{ext}} +\hat{\varphi})+{\phi(\hat{\boldsymbol\mu})}\cos({\phi_\mathrm{ext}} +\hat{\varphi})]\hat{{\sigma}}_a^x + \frac{1}{2}{E}_Z\hat{{\sigma}}_a^z\\
      &=-E_\mathrm{SO}{\phi(\hat{\boldsymbol\mu})}\cos({\phi_\mathrm{ext}}+\hat{\varphi})\hat{{\sigma}}_a^x-E_\mathrm{SO}\sin({\phi_\mathrm{ext}} +\hat{\varphi})\hat{{\sigma}}_a^x+\frac{1}{2}{E}_Z\hat{{\sigma}}_a^z.
      \label{eq40}
    \end{align}
    The last two terms include the free Hamiltonian of ASQ and the two-body interaction between ASQ and SCQ, which will be discussed in the next section. In terms of the transmon phase operator $\hat{\tilde{\varphi}}=\hat{\varphi}-\mathrm{arctan}(a\tan{\phi_\mathrm{ext}})$, the frist term in Eq.~\ref{eq40} reads
    \begin{align}
    \hat{H}_{\mathrm{int1}}&=    -E_\mathrm{SO}\phi(\hat{\boldsymbol\mu})\hat{{\sigma}}_a^x\left[\cos{\phi_\mathrm{ext}}\cos{\left(\hat{\tilde{\varphi}}+\arctan(a\tan\phi_\mathrm{ext})\right)}-\sin{\phi_\mathrm{ext}}\sin{\left(\hat{\tilde{\varphi}}+\arctan(a\tan\phi_\mathrm{ext})\right)}\right].
    \end{align}
    By the trigonometric relations
    \begin{align}
        \cos\left(  \arctan{x}\right)    =1/\sqrt{1+x^{2}} ,\quad\sin\left(  \arctan{x}\right)     =x/\sqrt{1+x^{2}} ,
        \label{eq16}
    \end{align}
    the above Hamiltonian can be simplified to
    \begin{equation}
       \hat{H}_{\mathrm{int1}}=    \frac{E_\mathrm{SO}\phi(\hat{\boldsymbol\mu})\hat{{\sigma}}_x}{S(\phi_\mathrm{ext})} \left[ \frac{(1+a)\sin{2\phi_\mathrm{ext}}}{2}\sin{\hat{\tilde{\varphi}}}+(\cos^2{\phi_\mathrm{ext}}-a\sin^2\phi_\mathrm{ext})\cos{\hat{\tilde{\varphi}}}\right].
    \end{equation}
    Expanding the trigonometric functions and retaining the first two orders, we obtain
    \begin{align}
    &\hat{H}_{\mathrm{int1}}=\frac{E_\mathrm{SO}\phi(\hat{\boldsymbol\mu})\hat{{\sigma}}_a^x}{S(\phi_\mathrm{ext})}\left[\frac{(1+a)\sin{2\phi_\mathrm{ext}}}{2} \hat{\tilde{\varphi}}+(\cos^2{\phi_\mathrm{ext}}-a\sin^2\phi_\mathrm{ext})( 1- \frac{\hat{\tilde{\varphi}}^{2}}{2})\right]
    \end{align}
    Substituting the quantized flux and phase difference in terms of the annihilation (creation) operators $\hat{m}^{(\dagger)}$ and $\hat{s}^{(\dagger)}$, the above expression reads
    \begin{align}
    \hat{H}_{\mathrm{int1}}=\frac{E_\mathrm{SO}\pi\Phi_\mathrm{YIG}(\hat{m}^{\dagger}+\hat{m})\hat{{\sigma}}_a^x}{S(\phi_\mathrm{ext})\Phi_0}\left\{\frac{(1+a)\sin{2\phi_\mathrm{ext}}{(2E_C)}^{1/4}}{2({E_{J}^\mathrm{sum}S(\phi_\mathrm{ext})})^{1/4}}(\hat{s}^\dagger+\hat{s}) +(\cos^2{\phi_\mathrm{ext}}-a\sin^2\phi_\mathrm{ext})[ 1-\frac{({2E_C})^{1/2}(\hat{s}^\dagger+\hat{s})^2}{2({E_{J}^\mathrm{sum}S(\phi_\mathrm{ext}))}^{1/2}}]\right\}.
    \label{eq46}
    \end{align}
    In the two-level subspace $\{\ket{g},\ket{e}\}$ of ASQ ($\hat{s}\rightarrow\hat{\sigma}^-_s,\hat{s}^\dagger\rightarrow\hat{\sigma}^+_s$), Eq.~\ref{eq46} becomes
    \begin{equation}
        \hat{H}_{\mathrm{int1}}=\hat{H}_\mathrm{thr}+\hat{H}_\mathrm{two1},
    \end{equation}
    where
    \begin{align}
        &\hat{H}_{\mathrm{thr}}/\hbar= G(\hat{\sigma}_s^-+\hat{\sigma}_s^+)(\hat{m}+\hat{m}^\dagger)\hat{\sigma}_a^x+ J
    \hat{\sigma}_s^+ \hat{\sigma}_s^-(\hat{m}+\hat{m}^\dagger)\hat{\sigma}_a^x,\label{eq19}\\ &\hat{H}_\mathrm{two1}/\hbar=g_1({m}\hat{\sigma}_a^++\hat{m}^\dagger\hat{\sigma}_a^-),\label{eq20}
    \end{align}
    are three-body and two-body interactions between magnon and ASQ, respectively, with the rotating-wave approximation applied to disregard fast rotating terms, like the terms with $\hat{m}\hat{\sigma}_a^-$ or $\hat{m}^\dagger\hat{\sigma}_a^+$ in Eq.~\ref{eq20}. Note that terms with $(\hat{\sigma}_s^-)^{2}$ or $(\hat{\sigma}_s^+)^2$ in Eq.~\ref{eq19} are identically zero in the qubit subspace and are therefore omitted.
    The coupling strength of the first term of three-body interaction is expressed as
    \begin{equation}
            G =\frac{E_\mathrm{SO}\sin2\phi_\mathrm{ext}(1+a)\pi\Phi_\mathrm{YIG}}{2\hbar\Phi_0S(\phi_{ext})^{5/4}}(\frac{2E_C}{E_{J}^{\mathrm{sum}}})^{1/4},
    \end{equation}
    while the second term of three-body interaction has a coupling strength
    \begin{equation}
        J =\frac{E_\mathrm{SO}(a\sin^2\phi_\mathrm{ext}-\cos^2\phi_\mathrm{ext})\pi\Phi_\mathrm{YIG}}{\hbar\Phi_0S(\phi_{ext})^{3/2}}(\frac{2E_C}{E_{J}^{\mathrm{sum}}})^{1/2}.
    \end{equation}
    The coupling strength of the two-body interaction between the magnon and the SCQ is
    \begin{equation}
        g_1 =-\frac{E_\mathrm{SO}(a\sin^2\phi_\mathrm{ext}-\cos^2\phi_\mathrm{ext})\pi\Phi_\mathrm{YIG}}{\hbar\Phi_0S(\phi_{ext})}[1-\frac{1}{2}(\frac{2E_C}{E_{J}^{\mathrm{sum}}S(\phi_{ext})})^{1/2}].
    \end{equation}

\begin{figure}
        \centering
        \includegraphics[width=0.6\linewidth]{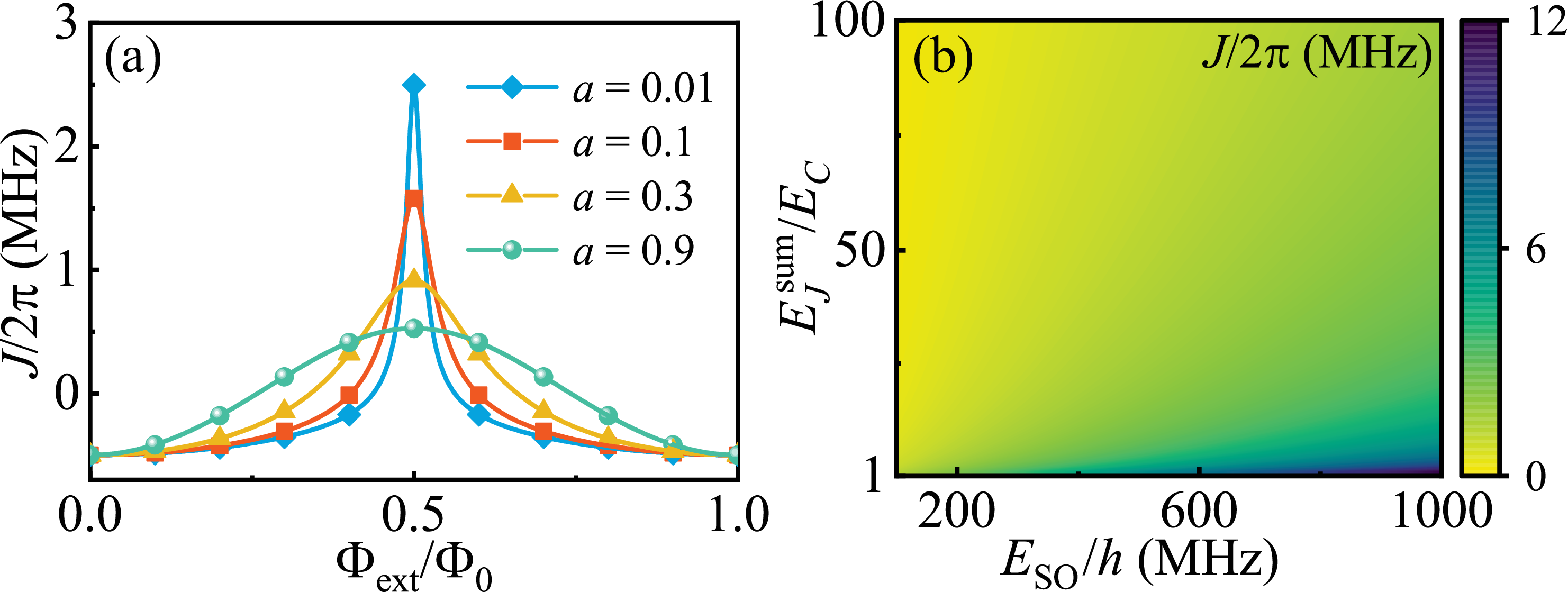}
        \caption{(a) Variation of the coupling strengths $J$ with
     the external flux $\Phi_\mathrm{ext}$ for different asymmetry $a$. (b) Contour map of the coupling strength $J$ versus the ratio $E_{J}^{\mathrm{sum}}/E_C$ and spin-dependent energy $E_\mathrm{SO}$ with $a = 0.1$.}
        \label{f4}
    \end{figure}

     As shown in Fig.~\ref{f4}(a), the coupling strength $J$ is plotted as functions of the external flux $\Phi_\mathrm{ext}$ with different asymmetry $a$. The coupling strength $J$ reaches its maximum value when $\Phi_\mathrm{ext}=0.5$ and its maximum value decreases as the asymmetry $a$ increases. According to the contour map shown in Fig.~\ref{f4}(b), $J$ is also proportional to $E_\mathrm{SO}$ and decreases with increasing $E_{J}^{\mathrm{sum}}/E_C$. Moreover, the coupling strength $J$ is more strongly influenced by $E_{J}^{\mathrm{sum}}/E_C$ due to $J\propto(E_C/E_{J}^{\mathrm{sum}})^{1/2}$ compared with the coupling strength $G$.
    \subsection{Two-body interaction between ASQ and SCQ}

    We now discuss the last two terms in Eq.~\ref{eq40}. In terms of the transmon phase operator $\hat{\tilde{\varphi}}$, these terms can be rewritten as
    \begin{align}
     \hat{H}_\mathrm{int2}&=-E_\mathrm{SO}\left[\sin\phi_\mathrm{ext}\cos{{(\hat{\tilde{\varphi}}+\arctan(a\tan\phi_\mathrm{ext}))}} +\cos\phi_\mathrm{ext}\sin{{\left(\hat{\tilde{\varphi}}+\arctan(a\tan\phi_\mathrm{ext})\right)}}\right]\hat{{\sigma}}_a^x+\frac{1}{2}{E}_Z\hat{{\sigma}}_a^z\nonumber\\
     &=\frac{E_\mathrm{SO}\hat{{\sigma}}_x}{S(\phi_{ext})}\left[ (a\sin^2\phi_\mathrm{ext}-\cos^2\phi_\mathrm{ext})\sin{\hat{\tilde{\varphi}}}-\frac{(1+a)\sin2\phi_\mathrm{ext}}{2}\cos{\hat{\tilde{\varphi}}}\right]+\frac{1}{2}{E}_Z\hat{{\sigma}}_a^z
     \label{eq24}
    \end{align}
    Similarly, expanding up to the second order in Eq.~\ref{eq24}, we have
    \begin{align}
     \hat{H}_\mathrm{int2}=\frac{E_\mathrm{SO}\hat{{\sigma}}_x}{S(\phi_{ext})}\left[ (a\sin^2\phi_\mathrm{ext}-\cos^2\phi_\mathrm{ext})\hat{\tilde{\varphi}}-\frac{(1+a)\sin2\phi_\mathrm{ext}}{2}(1-\frac{{\hat{\tilde{\varphi}}^2}}{2})\right]+\frac{1}{2}{E}_Z\hat{{\sigma}}_a^z.
    \end{align}
    Then substituting the quantized flux and phase difference, the Hamiltonian becomes
    \begin{align}
       \hat{H}_2=\hat{H}_\mathrm{ASQ}+\hat{H}_\mathrm{two2}.
    \end{align}
    Here
    \begin{align}
        \hat{H}_\mathrm{ASQ}=-E_\mathrm{SO}\frac{(1+a)\sin2\phi_\mathrm{ext}}{2S(\phi_{ext})}\hat{\sigma}_a^x+\frac{1}{2}{E}_Z\hat{{\sigma}}_a^z\simeq \frac{\hbar\omega_a}{2}\hat{\sigma}_a^z
    \end{align}
    are the free Hamiltonian of ASQ with the frequency of ASQ $\omega_a\simeq E_Z/\hbar$ for large Zeeman energy ($E_Z\gg E_\mathrm{SO}$). And the two-body interaction $\hat{H}_\mathrm{two2}$ between ASQ and SCQ can be expressed as
    \begin{align}
    \hat{H}_\mathrm{two2}=\hbar g_2(\hat{\sigma}_s^-\hat{\sigma}_a^++\hat{\sigma}_s^+\hat{\sigma}_a^-)+\hbar \bar{g}_2\hat{\sigma}_s^+\hat{\sigma}_s^-(\hat{\sigma}_a^++\hat{\sigma}_a^-) .\label{eq28}
    \end{align}
    The first term in Eq.~\ref{eq28} describes the coherent exchange between ASQ and SCQ, and its coupling strength is
    \begin{equation}
        g_2=\frac{E_\mathrm{SO}(a\sin^2\phi_\mathrm{ext}-\cos^2\phi_\mathrm{ext})}{S(\phi_\mathrm{ext})^{5/4}} (\frac{2E_C}{E_{J}^\mathrm{sum}})^{1/4}.
    \end{equation}
    The second term has a coupling strength
    \begin{equation}
     \bar{g}_2=\frac{E_\mathrm{SO}(1+a)\sin2\phi_\mathrm{ext}}{2S(\phi_\mathrm{ext})^{3/2}}(\frac{2E_C}{E_{J}^\mathrm{sum}})^{1/2} ,
    \end{equation}
    which represents the radiation-pressure type interaction.
    \subsection{Two-body interaction between the magnon and SCQ}
    According to the trigonometric relations in Eq.~\ref{eq16}, the Hamiltonian in Eq.~\ref{eq14} can be expressed as
    \begin{align}
    \hat{\tilde{H}}_\mathrm{SCQ}&=4E_{C}\hat{N}^{2} -E_{J}^\mathrm{sum}[\cos(\phi_\mathrm{ext}+\phi(\hat{\boldsymbol\mu}))\cos \hat{\varphi}+a\sin(\phi_\mathrm{ext}+\phi(\hat{\boldsymbol\mu}))\sin \hat{\varphi}] \label{eq31},
    \end{align}
    By expanding the trigonometric functions with $\phi(\hat{\boldsymbol\mu})\ll1$, we simplify the Hamiltonian as
    \begin{align}
    \hat{\tilde{H}}_\mathrm{SCQ}&=4E_{C}\hat{N}^{2} -E_{J}^\mathrm{sum}[\cos\phi_\mathrm{ext}\cos \hat{\varphi}-\phi(\hat{\boldsymbol\mu})\sin\phi_\mathrm{ext}\cos \hat{\varphi}+a\sin\phi_\mathrm{ext}\sin \hat{\varphi}+\phi(\hat{\boldsymbol\mu})a\cos\phi_\mathrm{ext}\sin \hat{\varphi}] \nonumber\\
    &=\hat{H}_\mathrm{ASQ}+\hat{H}_\mathrm{two3},\label{eq32}
    \end{align}
    where
    \begin{align}
        \hat{H}_\mathrm{SCQ}&=4E_{C}\hat{N}^{2} -E_{J}^\mathrm{sum}\cos\phi_\mathrm{ext}[\cos \hat{\varphi}+a\tan\phi_\mathrm{ext}\sin \hat{\varphi}]\nonumber\\
        &=\frac{\hbar\omega_s}{2} \hat{\sigma}_s^z,\\
        \hat{H}_\mathrm{two3}&=E_{J}^\mathrm{sum}\phi(\hat{\boldsymbol\mu})[\sin\phi_\mathrm{ext}\cos \hat{\varphi}-a\cos\phi_\mathrm{ext}\sin \hat{\varphi}],
    \end{align}
    denote the free Hamiltonian of SCQ and two-body interaction between magnon and SCQ. In terms of the transmon phase operator $\hat{\tilde{\varphi}}$, the interaction Hamiltonian becomes
    \begin{equation}
    \hat{H}_\mathrm{two3}=E_{J}^\mathrm{sum}\phi(\hat{\boldsymbol\mu})\left[
    \sin\phi_\mathrm{ext} \cos\left(
    \hat{\tilde{\varphi}} + \arctan\left(a \tan\phi_\mathrm{ext}\right)
    \right)
    - a \cos\phi_\mathrm{ext} \sin\left(
    \hat{\tilde{\varphi}} + \arctan\left(a \tan\phi_\mathrm{ext}\right)
    \right)
    \right]
    \end{equation}
    and can be further simplified to
    \begin{equation}
        \hat{H}_\mathrm{two3}=\frac{E_{J}^\mathrm{sum}\phi(\hat{\boldsymbol\mu})}{S(\phi_\mathrm{ext})}\left[ \frac{\sin 2\phi_\mathrm{ext}}{2}  (1 - a^2) \cos \hat{\tilde{\varphi}} - a \sin \hat{\tilde{\varphi}}  \right].
    \end{equation}
    Then substituting the quantized magnetic flux and phase expressed by bosonic operators and Pauli operators, respectively, the above expression reads
    \begin{equation}   \hat{H}_\mathrm{two3}=g_3(\hat{m}\hat{\sigma}_s^++\hat{m}^\dagger\hat{\sigma}_s^-)+\bar{g}_3\hat{\sigma}_s^+\hat{\sigma}_s^-(\hat{m}^\dagger+\hat{m}).\label{eq37}
    \end{equation}
    The first term represents the coherent exchange between the magnon and the SCQ, whose coupling strength is expressed as
    \begin{equation}
     g_3=-\frac{a(E_{J}^\mathrm{sum})^{3/4}({2E_C})^{1/4}\pi\Phi_\mathrm{YIG}}{S(\phi_\mathrm{ext})^{5/4}\Phi_0}
    \end{equation}
    The second term denotes radiation-pressure type interaction with
    \begin{equation}
     \bar{g}_3= -   \frac{(1 - a^2)\sin 2\phi_\mathrm{ext}(E_{J}^\mathrm{sum})^{1/2}({2E_C})^{1/2}\pi\Phi_\mathrm{YIG}}{2S(\phi_\mathrm{ext})^{3/2}\Phi_0}.
    \end{equation}
    \subsection{Negligibility of off-resonant two-body interactions and dynamical verification}
    Based on the results of the derivation in the previous text, the total Hamiltonian of the system can be expressed as
    \begin{align}
        \hat{H}_\mathrm{tot}=\hat{H}_M+\hat{H}_\mathrm{ASQ}+\hat{H}_\mathrm{SCQ}+\hat{H}_\mathrm{thr}+\hat{H}_\mathrm{two},
    \end{align}
    where $\hat{H}_M$, $\hat{H}_\mathrm{ASQ}$ and $\hat{H}_\mathrm{SCQ}$ denote the free Hamiltonian of magnon, ASQ and SCQ, respectively. The three-body interaction can be expressed
    \begin{equation}
        \hat{H}_{\mathrm{thr}}= \hbar G(\hat{\sigma}_s^-+\hat{\sigma}_s^+)(\hat{m}+\hat{m}^\dagger)\hat{\sigma}_a^x+ \hbar J
    \hat{\sigma}_s^+ \hat{\sigma}_s^-(\hat{m}\hat{\sigma}_a^++\hat{m}^\dagger\hat{\sigma}_a^-),
    \label{eq70}
    \end{equation}
    which concludes the linear and nonlinear three-body interaction. And $\hat{H}_\mathrm{two}$ contains all the two-body interaction Hamiltonians, which reads
    \begin{align}
    \hat{H}_\mathrm{two}&=\hat{H}_\mathrm{two1}+\hat{H}_\mathrm{two2}+\hat{H}_\mathrm{two3},\nonumber\\
    \hat{H}_\mathrm{two1}&=\hbar g_1({m}\hat{\sigma}_a^++\hat{m}^\dagger\hat{\sigma}_a^-),\nonumber\\
    \hat{H}_\mathrm{two2}&=\hbar g_2(\hat{\sigma}_s^-\hat{\sigma}_a^++\hat{\sigma}_s^+\hat{\sigma}_a^-)+\hbar \bar{g}_2\hat{\sigma}_s^+\hat{\sigma}_s^-(\hat{\sigma}_a^++\hat{\sigma}_a^-) ,\nonumber\\
    \hat{H}_\mathrm{two3}&=g_3(\hat{m}\hat{\sigma}_s^++\hat{m}^\dagger\hat{\sigma}_s^-)+\bar{g}_3\hat{\sigma}_s^+\hat{\sigma}_s^-(\hat{m}^\dagger+\hat{m}).
    \end{align}

\begin{figure}[tp]
        \centering
    \includegraphics[width=0.6\columnwidth]{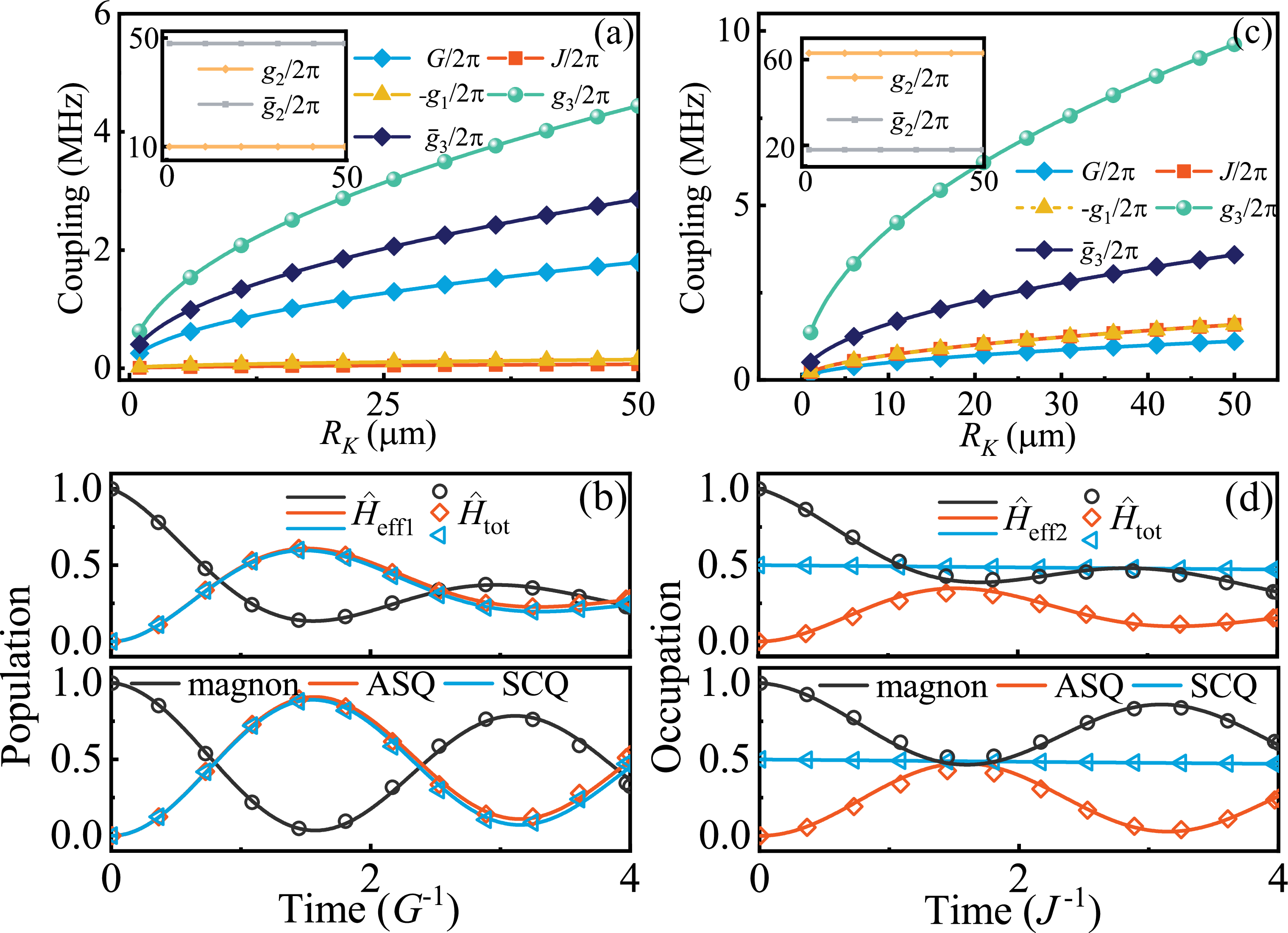}
    \caption{\label{figflux} All coupling strengths as functions of magnetic sphere radius $R_K$ for resonance condition (a) $\omega_a=\omega_m+\omega_s$ ($\omega_m=\omega_a+\omega_s$) and (c) $\omega_m=\omega_a$. The three-body interaction induce dynamical evolution under the resonance condition (c) $\omega_m=\omega_a+\omega_s$ and (d) $\omega_m=\omega_a$. The dissipation rates are reduced as $\kappa_m/2\pi=0.1~\mathrm{MHz}$ and $\gamma_a/2\pi=0.1~\mathrm{MHz}$ in the lower panel.}
    \label{f5}
    \end{figure}

    Next, we discuss the influence of the two-body interaction  $\hat{H}_\mathrm{two}$ on cases dominated by three-body interaction $\hat{H}_\mathrm{thr}$. When the resonance condition $\omega_m=\omega_a+\omega_s$ (or $\omega_a=\omega_m+\omega_s$) is adopted, the second term in Eq.~\ref{eq70} is far off-resonance and can be neglected, resulting in the three-body interaction becoming $\hat{m} \hat{\sigma}_s^+\hat{\sigma}_a^++\hat{m}^\dagger \hat{\sigma}_s^-\hat{\sigma}_a^-$ (or $\hat{m}^\dagger\hat{\sigma}^+_s\hat{\sigma}^-_a+\hat{m} \hat{\sigma}^-_s\hat{\sigma}^+_a$) under the rotating wave approximation. As shown in Fig.~\ref{f5}(a), all coupling strengths of three-body and two-body interaction are plotted as functions of the radius $R_K$. It is obvious that, except for the two-body coupling strength between the two qubits, all other coupling strengths increase with the increase of the magnetic sphere radius $R_K$ due to $\Phi_\mathrm{ext}\propto R^{1/2}$. Here, we consider the typical transmon parameter $E_{J}^{\mathrm{sum}}/h=10\mathrm{GHz}$ and $E_C/h=200\mathrm{MHz}$, external flux $\Phi_\mathrm{ext}/\Phi_0=0.35$ and $a=0.3$, which results in the frequency of the SCQ $\omega_s/2\pi=2.5\mathrm{GHZ}$. The frequency of magnon and ASQ are chosen as $\omega_m/2\pi=7.5\mathrm{GHZ}$ ($\omega_m/2\pi=12.5\mathrm{GHZ}$) and $\omega_a/2\pi=10\mathrm{GHZ}$ to satisfy $\omega_a=\omega_m+\omega_s$ ($\omega_m=\omega_a+\omega_s$), which can be realized by adjusting the strengths of magnetic field $B_Z$ applied to the quantum dot junction and $B_K$ applied to the YIG sphere. The coupling strength $\bar{g}_2$ of the radiation-pressure type interaction between the two qubits is about $48\times2\pi\mathrm{MHz}$, which is larger than that of other interactions. However, the frequency of ASQ $\omega_a$ is much greater than $\bar{g}_2$ and the detuning $\Delta_q=7.5\times2\pi\mathrm{GHz}$ between the two qubits is also much greater than the coupling strength ${g}_2$. So the rapidly oscillating term $\hat{H}_\mathrm{two2}$ can be neglected. Similarly, $\hat{H}_\mathrm{two1}$ and $\hat{H}_\mathrm{two3}$ also oscillate rapidly and can be ignored. Therefore, the influence of all two-body interactions can basically be neglected. As a result, for $\omega_m=\omega_a+\omega_s$, the effective Hamiltonian can be expressed as
    \begin{equation}
    \hat{H}_\mathrm{eff}=\hat{H}_M+\hat{H}_\mathrm{ASQ}+\hat{H}_\mathrm{SCQ}+\hbar G(\hat{m}^\dagger\hat{\sigma}^+_s\hat{\sigma}^-_a+\hat{m} \hat{\sigma}^-_s\hat{\sigma}^+_a), \label{eq72}
    \end{equation}
     while for $\omega_m=\omega_a+\omega_s$, the effective Hamiltonian becomes
    \begin{equation}
    \hat{H}_\mathrm{eff1}=\hat{H}_M+\hat{H}_\mathrm{ASQ}+\hat{H}_\mathrm{SCQ}+\hbar G(\hat{m} \hat{\sigma}^+_s\hat{\sigma}^+_a+\hat{m}^\dagger \hat{\sigma}^-_s\hat{\sigma}^-_a).  \label{eq73}
    \end{equation}
    These two effective Hamiltonians, Eqs.~\ref{eq72} and \ref{eq73}, correspond to the two distinct quantum processes highlighted in the main text: (i) the excitation of a magnon and the SCQ upon the annihilation of the ASQ, and the reverse process; (ii) and the joint excitation of both qubits upon the annihilation of a magnon, and the reverse process. As shown in Fig.~\ref{f5} (b), we plot the population evolution of the system under the interaction $G(\hat{m} \hat{\sigma}^+_s\hat{\sigma}^+_a+\hat{m}^\dagger \hat{\sigma}^-_s\hat{\sigma}^-_a)$ with the initial state $\ket{1,\downarrow,g}$. Almost-perfect joint excitation of two qubits by magnons is observed, whereby the excitation of one qubit invariably implies the excitation of the other. This figure also exhibits
    excellent agreement between the dynamics generated by $\hat{H}'_\mathrm{eff}$ and $\hat{H}_\mathrm{tot}$, which confirms that neglecting these far off-resonance two-body interactions is justified.

    Then, we select the resonance condition $\omega_m=\omega_a$. The three-body interaction becomes $\hat{H}_\mathrm{thr}=\hbar J\hat{\sigma}_s^+\hat{\sigma}_s^-(\hat{m} \hat{\sigma}_a^++\hat{m}^\dagger \hat{\sigma}_a^-)$ with the neglected first term in Eq.~\ref{eq70} under the rotating wave approximation. Choosing $\Phi_\mathrm{ext}/\Phi_0=0.48$ and $a=0.1$, the frequency $\omega_s$ of SCQ is about $2\pi\times1.3\mathrm{GHZ}$. The frequency of magnon and ASQ are assumed as $\omega_m/2\pi=\omega_a/2\pi=10\mathrm{GHZ}$. We can find that the coupling strength $J$ of the second term of three-body interaction is approximately equal to the opposite of coupling strength $g_1$ of two-body interaction $\hat{H}_\mathrm{two1}$ between the magnon and ASQ with $g/2\pi = -1.2 \mathrm{MHz}$ when $R_K=30\mu m$ in Fig.~\ref{f5}(c). Moreover, the detuning $\Delta_q/2\pi=8.7\mathrm{GHz}$ between the two qubits and the detuning $\Delta_m/2\pi=8.7\mathrm{GHz}$ between the magnon and SCQ are much larger than the coupling strengths $g_2$ and $g_3$, respectively. So, the remaining far-resonant two-body interactions $\hat{H}_\mathrm{two2}$ and $\hat{H}_\mathrm{two3}$ oscillate rapidly and can be neglected. As a result, the interaction of the system can be written as $\hbar J(\hat{\sigma}_s^+ \hat{\sigma}_s^--1)(\hat{m}\hat{\sigma}_a^++\hat{m}^\dagger\hat{\sigma}_a^-)$, and effective Hamiltonian can be expressed as
    \begin{equation}
    \hat{H}_\mathrm{eff2}=\hat{H}_M+\hat{H}_\mathrm{ASQ}+\hat{H}_\mathrm{SCQ}+\hbar J(\hat{\sigma}_s^+ \hat{\sigma}_s^--1)(\hat{m}\hat{\sigma}_a^++\hat{m}^\dagger\hat{\sigma}_a^-),
    \end{equation}
     As shown in Fig.~\ref{f5}(d), the SCQ acts as a quantum switch to control the coherent exchange between magnon and ASQ with the initial state \( (|1,\downarrow,g\rangle + |1,\downarrow,e\rangle)/\sqrt{2} \). When the SCQ is in the ground state $|g\rangle$, there is a coherent exchange between the magnon and the ASQ; otherwise, no such coherent exchange exists.
    The dynamical evolution shown in Fig.~\ref{f5}(d) can also verify that these far off-resonance interactions can be safely neglected.
    \section{\label{secIII}Exponentially enhanced coupling strength}
\begin{figure}[tp]
        \centering
    \includegraphics[width=0.6\columnwidth]{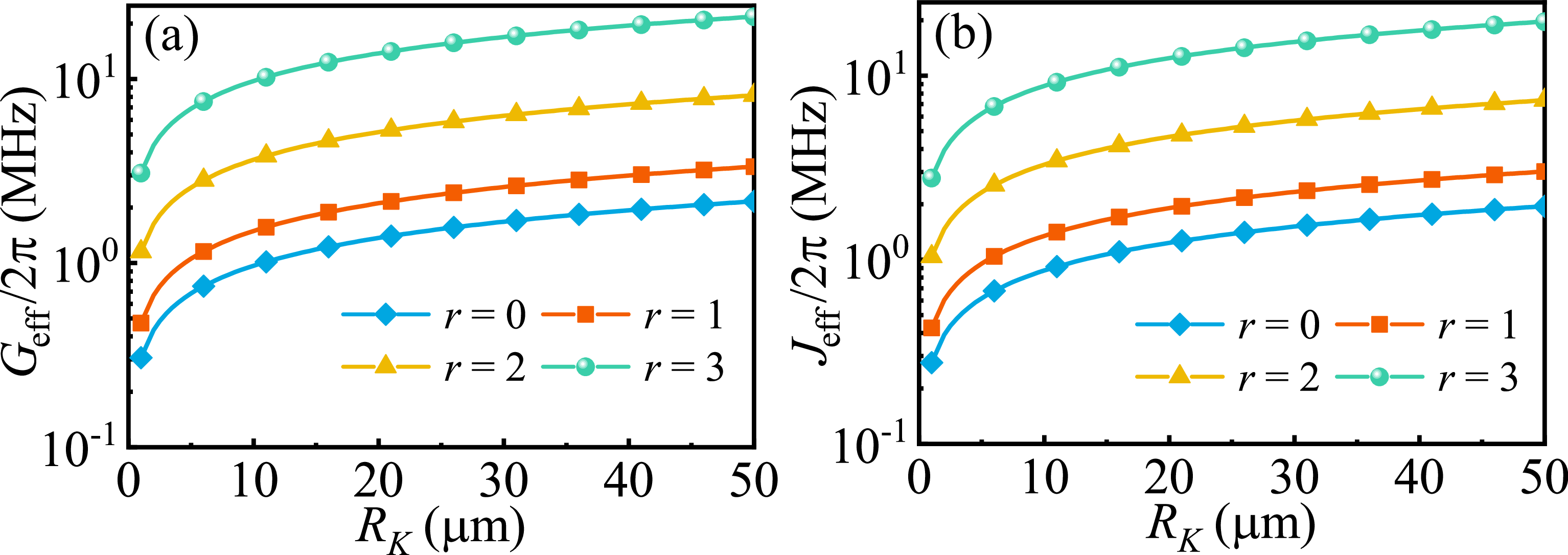}
    \caption{ (a) and (b) show the
     variation of the coupling strengths $G$ and $J$ with radius $R_K$ for different squeezing parameter $r$.}
    \label{f6}
    \end{figure}

    To enhance the coupling strength, we consider the magnon-Kerr effect arising from the anisotropy of the YIG sphere~\cite{PhysRevB.94.224410}. Then the Hamiltonian of the system can be expressed as
    \begin{align}
    \hat{H'}_\mathrm{sys}/\hbar=\omega_m\hat{m}^\dagger\hat{m}+\frac{{\omega_a}}{2}\hat{\sigma}^z_a+\frac{{\omega_s}}{2}\hat{\sigma}^z_s+G(\hat{m}^\dagger\hat{\sigma}^-_s\hat{\sigma}^-_a+\hat{m} \hat{\sigma}^+_s\hat{\sigma}^+_a)-K\hat{m}^{\dagger}\hat{m}\hat{m}^{\dagger}\hat{m},
    \end{align}
    where $K=\mu_0\gamma_0^2K_{an}/(M_s^2V_K)$ denotes the Kerr coefficient with the first-order anisotropy constant $K_{an}$ and the saturation magnetization $M_s$. The Kerr coefficient is inversely proportional to the volume of the YIG sphere $V_K$. The Kerr effect based on the anisotropy of YIG spheres is extremely weak. Thus, we apply microwave drive to enhance the Kerr effect. The Hamiltonia with microwave drive is written as
    \begin{align}
    \hat{H}_{sys}/\hbar=\omega_m\hat{m}^\dagger\hat{m}+\frac{{\omega_a}}{2}\hat{\sigma}^z_a+\frac{{\omega_s}}{2}\hat{\sigma}^z_s+G(\hat{m}^\dagger\hat{\sigma}^-_s\hat{\sigma}^-_a+\hat{m} \hat{\sigma}^+_s\hat{\sigma}^+_a)-K\hat{m}^{\dagger}\hat{m}\hat{m}^{\dagger}\hat{m}+\Omega_d(\hat{m}^{\dagger}e^{-i\omega_dt}+\hat{m} e^{i\omega_dt})    ,
    \end{align}
    where $\Omega_d$ represents the drive strength, and $\omega_d$ represents the drive frequency. In the rotating frame with respect to $    V=\omega_d\hat{m}^\dagger\hat{m}+{\omega_d}\hat{\sigma}^z_a/4+{\omega_ d}\hat{\sigma}^z_a/4$, the Hamiltonian is simplified to
    \begin{equation}
    \hat{H}/\hbar=\Delta_m\hat{m}^\dagger\hat{m}+\frac{{\Delta_a}}{2}\hat{\sigma}^z_a+\frac{{\Delta_s}}{2}\hat{\sigma}^z_s+G(\hat{m}^\dagger\hat{\sigma}^-_s\hat{\sigma}^-_a+\hat{m} \hat{\sigma}^+_s\hat{\sigma}^+_a)-K\hat{m}^{\dagger}\hat{m}\hat{m}^{\dagger}\hat{m}+\Omega_d(\hat{m}^{\dagger}+\hat{m} ) ,
    \end{equation}
    where $\Delta_m=\omega_m-\omega_d$, $\Delta_a=\omega_a-\omega_d/2$ and $\Delta_s=\omega_s-\omega_d/2$. Using the Heisenberg dynamical equation $\dot{\hat{A}} = i\left[\hat{H}, \hat{A}\right]$,  the dynamical equation for the magnon is written as
    \begin{equation}
        \dot{\hat{m}}/\hbar = -i (\Delta_m-K) \hat{m} - i G (\hat{\sigma}^-_a  \hat{\sigma}^-_s) + 2i K \hat{m}^\dagger \hat{m} \hat{m} - i \Omega_d
    \end{equation}
    Assuming the microwave drive is strong, the operator can be expressed as its expected value plus its associated fluctuation $\hat{A} \to \langle \hat{A} \rangle + \hat{A} $. Neglecting higher-order fluctuation terms in the strong driving regime, the magnon's dynamical equation admits a simplification to
    \begin{equation}
        \dot{\hat{m}}/\hbar = -i (\Delta_m-K) \hat{m} - i G (\hat{\sigma}^-_a  \hat{\sigma}^-_s) +4i K \langle \hat{m} \rangle^2 \hat{m} + 2i K \langle \hat{m} \rangle^2 \hat{m}^\dagger.
    \end{equation}
    Then the linearized Hamiltonian can be written as
    \begin{equation}
    \hat{H}/\hbar=\widetilde{\Delta}_m\hat{m}^\dagger\hat{m}+\frac{{\Delta_a}}{2}\hat{\sigma}^z_a+\frac{{\Delta_s}}{2}\hat{\sigma}^z_s+G(\hat{m}^\dagger\hat{\sigma}^-_s\hat{\sigma}^-_a+\hat{m} \hat{\sigma}^+_s\hat{\sigma}^+_a)- \frac{K_d}{2} (\hat{m}^{\dagger 2} + \hat{m}^2)     ,
    \label{eq75}
    \end{equation}
    where detuing $\widetilde{\Delta}_m = \Delta_m -K- 4K \langle \hat{m}  \rangle^2$ and enhanced Kerr coefficient $K_d = 2K \langle \hat{m} \rangle^2$. Using the Bogoliubov transformation $\hat{m} = \hat{s}_K \cosh r - \hat{s}^\dagger_K \sinh r$, with $\tanh(2r) = K_d/\widetilde{\Delta}_m$~\cite{lemonde2016enhanced,doi:10.1126/science.aaw2884}, the Hamiltonian $\hat{H}_{\mathrm{sys}}$ in Eq.~\ref{eq75} can be expressed as
    \begin{equation}
        \hat{H}^\mathrm{eff}_\mathrm{sys}/\hbar={\Delta}_m^\mathrm{eff}\hat{m}^\dagger\hat{m}+\frac{{\Delta_a}}{2}\hat{\sigma}^z_a+\frac{{\Delta_s}}{2}\hat{\sigma}^z_s+G_\mathrm{eff}(\hat{m}^\dagger\hat{\sigma}^-_s\hat{\sigma}^-_a+\hat{m} \hat{\sigma}^+_s\hat{\sigma}^+_a),
    \end{equation}
    where $\Delta_m^\mathrm{eff}=\widetilde{\Delta}_m/\cosh{2r}$ and $G_\mathrm{eff}=G\cosh{r}$. Naturally, the coupling strength $G_\mathrm{eff}$ increases exponentially with the squeeze parameter $r$. Similarly, the coupling strength $J$ can also be enhanced exponentially. As shown in Figs.~\ref{f6}(a) and ~\ref{f6}(b), we plot the variation of the effective coupling strengths $G_\mathrm{eff}$ and $J_\mathrm{eff}$ with radius $R$ for different the squeeze parameter $r$.

    \section{\label{secIV} Collapse-revival phenomena and entanglement redistribution}
    \subsection{Analytical solution for the collapse-revival dynamics}
    The main text briefly discussed the collapse-revival phenomena when the magnon is prepared in a coherent state $|\alpha\rangle$ under the resonance condition $\omega_a = \omega_m + \omega_s$. Here, we provide a more detailed theoretical derivation and parameter analysis to supplement these findings. The collapse-revival phenomena is explored by three-body interaction $G(\hat{m}^\dagger\hat{\sigma}^+_s\hat{\sigma}^-_a+\hat{m} \hat{\sigma}^-_s\hat{\sigma}^+_a)$, where the effective Hamiltonian of the system can be expressed as
    \begin{equation}   \hat{H}_\mathrm{eff}=\hat{H}_M+\hat{H}_\mathrm{ASQ}+\hat{H}_\mathrm{SCQ}+\hbar G(\hat{m}^\dagger\hat{\sigma}^+_s\hat{\sigma}^-_a+\hat{m} \hat{\sigma}^-_s\hat{\sigma}^+_a)
    \end{equation}
    We assume that the initial state of the system is the excited state of ASQ, the ground state of SCQ, and the coherent state of the magnon:
    \begin{equation}
     |\psi(0)\rangle =|\alpha\rangle\otimes |\!\uparrow\rangle\otimes|g\rangle
    \end{equation}
    where the coherent state is $|\alpha\rangle = e^{-|\alpha|^2/2} \sum_{n = 0}^{\infty} \frac{\alpha^n}{\sqrt{n!}} |n\rangle$ with the average magnon number $\bar{n} = |\alpha|^2$. The action of the Hamiltonian $H_{\mathrm{eff}}$ is confined to the subspace spanned by  $\{|n+1,\downarrow, e\rangle, |n,\uparrow,g\rangle\}$. Therefore, the time-evolved state can be expressed as
    \begin{equation}
        |\psi(t)\rangle = \sum_{n = 0}^{\infty} \left[ C_{n + 1,\downarrow,e}(t)|n + 1,\downarrow,e\rangle + C_{n,\uparrow,g}(t)|n,\uparrow,g\rangle\right]e^{-i(n+1/2)\omega_mt}.
    \end{equation}
    Substituting into the Schr\"{o}dinger equation $i\hbar \frac{d}{dt} |\psi\rangle = H_{\mathrm{eff}} |\psi\rangle$, we obtain the following coupled differential equations:
    \begin{equation}
    \begin{cases}
    i\dot{C}_{n + 1,\downarrow,e} = G\sqrt{n + 1} C_{n, \uparrow,g} \\
    i\dot{C}_{n,\uparrow,g} = G\sqrt{n + 1} C_{n + 1,\downarrow,e}
    \end{cases}
    \end{equation}
    Solving these equations, we get
    \begin{align}
        C_{n,\uparrow,g}(t) = \cos(\Omega_n t) \cdot e^{-|\alpha|^2 / 2} \frac{\alpha^n}{\sqrt{n!}},\quad C_{n + 1,\downarrow,e}(t) = -i \sin(\Omega_n t) \cdot e^{-|\alpha|^2 / 2} \frac{\alpha^{n}}{\sqrt{n!}},
    \end{align}
    where \(\Omega_n = G\sqrt{n + 1}\) is the Rabi frequency dependent on $n$.

\begin{figure}
        \centering
        \includegraphics[width=0.6\linewidth]{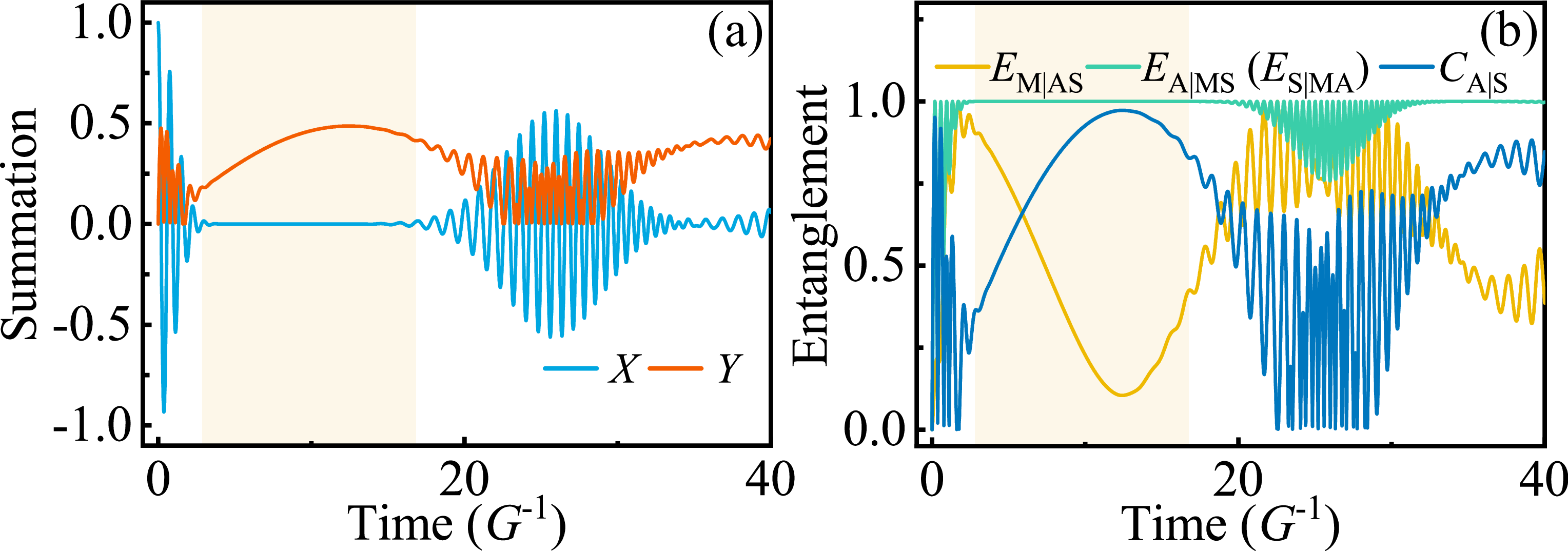}
        \caption{(a) The evolution of series summation $X$ and $Y$. (b) Entanglement evolution calculated by von Neumann entropy and concurrence analytically.}
        \label{fg7}
    \end{figure}

    The population of the ASQ $|\!\uparrow\rangle$ state is
    \begin{equation}
     P_\uparrow(t) = \sum_{n = 0}^{\infty} \left| C_{n,\uparrow,g}(t) \right|^2 = \sum_{n = 0}^{\infty} \left| \cos\left( \Omega_n t \right) \right|^2 \cdot e^{-|\alpha|^2 } \frac{\alpha^{2n}}{{n!}},
     \label{eq75}
    \end{equation}
    with $C_n=e^{-|\alpha|^2 } \frac{\alpha^{2n}}{{n!}}$.
    Using the identity $\cos^2{x}=1/2[1+\cos(2x)]$, Eq.~\ref{eq75} becomes
    \begin{equation}
     P_\uparrow(t) = \frac{1}{2} + \frac{1}{2} \sum_{n = 0}^{\infty} \cos\left( 2 \Omega_n t \right)\cdot e^{-|\alpha|^2 } \frac{\alpha^{2n}}{{n!}}.
    \end{equation}
    Similarly, the population of the SCQ $| e\rangle$ state can be expressed as
    \begin{align}
       P_e(t)& =\sum_{n = 0}^{\infty} \left| \sin\left( \Omega_n t \right) \right|^2 \cdot e^{-|\alpha|^2 } \frac{\alpha^{2n}}{{n!}} \\
     &=\frac{1}{2} - \frac{1}{2} \sum_{n = 0}^{\infty} \cos\left( 2 \Omega_n t  \right)\cdot e^{-|\alpha|^2 } \frac{\alpha^{2n}}{{n!}}.
    \end{align}
    It is obvious that the evolution of populations is determined by the series summation $X=\sum_{n = 0}^{\infty}C_n\cdot \cos\left( 2  \Omega_nt \right)$ with $C_n=e^{-|\alpha|^2 } \frac{\alpha^{2n}}{{n!}}$. As shown in Fig.~\ref{fg7} (a), when these components cancel each other out due to random phases, the series summation $X$ remains zero; while their superposition reaches a maximum when these components achieve phase synchronization. This result leads to the population oscillations of the two qubits exhibiting synchronized collapse and revival.

    \subsection{Logarithmic negativity}
    Logarithmic negativity $\mathcal{E}$~\cite{PhysRevA.54.1838} is a prominent measure for entanglement. It is defined as
    \begin{equation}
        \mathcal{E}= \log_2 \|\rho^{T}\|_1
    \end{equation}
    where $T$ denotes a partial transpose and $\|\cdot\|_1$ is the trace norm.
    For the three-body hybrid system, the entanglement can be expressed as
    \begin{align}
        \mathcal{E}_{i|jk}=\log_2 \|\rho^{T_i}\|_1,\quad\mathcal{E}_{i|j}=\log_2 \|\rho^{T_i}_{ij}\|_1,\quad\mathcal{E}_{i|k}=\log_2 \|\rho^{T_i}_{ik}\|_1.
    \end{align}
    Here, the reduced density matrix $\rho_{ij}$ ($\rho_{ik}$) can be obtained by taking the partial trace of the density matrix $\rho$.
    Equivalently, it can be expressed via the negativity \( \mathcal{N}(\rho) \) (quantifying violation of the positive partial transpose criterion) as \( E_\mathcal{N}(\rho) = \log_2[1 + 2\mathcal{N}(\rho)] \).
    \subsection{Von Neumann entropy and concurrence}
    For the entanglement between one subsystem and the combined system of the other two subsystems, we can also characterize it using the von Neumann entropy~\cite{PhysRevA.65.032314}. Firstly, we compute the reduced density matrix of the two qubits by taking the partial trace over the magnon degree of freedom:
    \begin{equation}
    \rho_q(t) = \mathrm{Tr}_{\mathrm{m}}\left( \ket{\psi(t)}\bra{\psi(t)} \right)
    \end{equation}
    The partial trace is calculated by summing over the magnon Fock states.  Expanding the total density matrix, we have
    \begin{align}
        \ket{\psi(t)}\bra{\psi(t)} &= \sum_{n,m} \left( C_{n + 1,\downarrow,e}(t)|n + 1,\downarrow,e\rangle + C_{n,\uparrow,g}(t)|n,\uparrow,g\rangle\right) \left( C_{m + 1,\downarrow,e}^*(t)\bra{m + 1,\downarrow,e} + C_{m,\uparrow,g}^*(t)\bra{m,\uparrow,g}\right) \nonumber \\
        & =\sum_{n,m} C_{n + 1,\downarrow,e}C_{m + 1,\downarrow,e}^*\ket{n + 1,\downarrow,e}\bra{m + 1,\downarrow,e}+\sum_{n,m} C_{n,\uparrow,g }C_{m,\uparrow,g }^*\ket{n,\uparrow,g }\bra{m,\uparrow,g}
        \nonumber\\
        &+ C_{n + 1,\downarrow,e}C_{m,\uparrow,g}^*\ket{n + 1,\downarrow,e}\bra{m,\uparrow,g}+ C_{n,\uparrow,g}C_{m + 1,\downarrow,e }^*\ket{n,\uparrow,g }\bra{m + 1,\downarrow,e}.
    \end{align}

    Taking a partial trace over the magnon subsystem involves evaluating the trace of each term. For any term with magnon components $\ket{n}\bra{m}$, the trace operation yields $\langle n \vert m \rangle = \delta_{n,m}$. Processing each term step-by-step, we have
    \begin{align}
    &\mathrm{Tr}_{m} \left( \sum_{n,m} C_{n+1,\downarrow,e}C_{m+1,\downarrow,e}^*\ket{n+1,\downarrow,e}\bra{m+1,\downarrow,e} \right)
    = \sum_{n=0}^{\infty}  C_{n+1,\downarrow,e} C_{n+1,\downarrow,e}^* \ket{\downarrow,e}\bra{\downarrow,e} = \sum_{n=0}^{\infty}  \vert C_{n+1,\downarrow,e} \vert^2 \ket{\downarrow,e}\bra{\downarrow,e}
    \\
    &\mathrm{Tr}_{m} \left( \sum_{n,m} C_{n,\uparrow,g }C_{m,\uparrow,g}^*\ket{n,\uparrow,g }\bra{m,\uparrow,g } \right)
    = \sum_{n=0}^{\infty}  C_{n,\uparrow,g } C_{n,\uparrow,g }^* \ket{\uparrow,g}\bra{\uparrow,g} = \sum_{n=0}^{\infty}  \vert C_{n,\uparrow,g } \vert^2 \ket{\uparrow,g}\bra{\uparrow,g,}
    \\
    &\mathrm{Tr}_{m} \left( \sum_{n,m} C_{n+1,\downarrow,e}C_{m,\uparrow,g}^*\ket{n+1,\downarrow,e }\bra{m,\uparrow,g } \right)
    = \sum_{n=1}^{\infty} C_{\downarrow,e,n  } C_{n,\uparrow,g }^* \ket{\downarrow,e}\bra{\uparrow,g} \\
    &\mathrm{Tr}_{m} \left( \sum_{n,m} C_{n,\uparrow,g }C_{m+1,\downarrow,e}^*\ket{n,\uparrow,g }\bra{m+1,\downarrow,e} \right)
    = \sum_{n=1}^{\infty} C_{n,\uparrow,g } C_{\downarrow,e,n }^* \ket{\uparrow,g}\bra{\downarrow,e}.
    \end{align}

    Combining all terms, the reduced density matrix of two qubits is
    \begin{align}
       \rho_q(t) =& \left( \sum_{n=0}^{\infty} \vert C_{n+1,\downarrow,e} \vert^2  \right) \ket{\downarrow,e}\bra{\downarrow,e}+ \left( \sum_{n=0}^{\infty} \vert C_{n,\uparrow,g } \vert^2  \right) \ket{\uparrow,g}\bra{\uparrow,g,}\nonumber \\
    &+ \left( \sum_{n=1}^{\infty}C_{\downarrow,e,n  } C_{n,\uparrow,g }^* \right) \ket{\downarrow,e}\bra{\uparrow,g}  + \left( \sum_{n=1}^{\infty}C_{n,\uparrow,g } C_{\downarrow,e,n }^* \right) \ket{\uparrow,g}\bra{\downarrow,e}
    \end{align}
    We can express the density matrix $\rho_q(t)$ in matrix form as follows:
    \begin{equation}
    \rho_q(t) = \begin{pmatrix}0&0&0&0\\
    0&\sum_{n=0}^{\infty} \vert C_{n+1,\downarrow,e} \vert^2 & \sum_{n=1}^{\infty} C_{\downarrow,e,n} C_{n,\uparrow,g}^*&0 \\
    0&\sum_{n=1}^{\infty} C_{n,\uparrow,g} C_{\downarrow,e,n}^* & \sum_{n=0}^{\infty} \vert C_{n,\uparrow,g} \vert^2&0\\
    0&0&0&0
    \end{pmatrix}.
    \end{equation}
    Here, the summations in the matrix elements of \(\rho_q(t)\) are expressed as
    \begin{align}
    &\sum_{n=1}^{\infty} C_{\downarrow,e,n} C_{n,\uparrow,g}^*=\sum_{n=1}^{\infty}i\cos(\Omega_n t)\sin(\Omega_{n-1} t) \cdot e^{-|\alpha|^2 } \frac{\alpha^{2n-1}}{{(n-1)!}\sqrt{n}},\\
    &\sum_{n=1}^{\infty} C_{n,\uparrow,g} C_{\downarrow,e,n}^*=\sum_{n=1}^{\infty}-i\cos(\Omega_n t)\sin(\Omega_{n-1} t) \cdot e^{-|\alpha|^2 } \frac{\alpha^{2n-1}}{{(n-1)!}\sqrt{n}}.\\
    \end{align}
    We diagonalize the density matrix to find its eigenvalues. The characteristic equation is
    \begin{align}
     \mathrm{det}
        \begin{pmatrix}
    \sum_{n=0}^{\infty} \vert C_{n+1,\downarrow,e} \vert^2-\lambda & \sum_{n=1}^{\infty} C_{\downarrow,e,n} C_{n,\uparrow,g}^* \\
    \sum_{n=1}^{\infty} C_{n,\uparrow,g} C_{\downarrow,e,n}^* & \sum_{n=0}^{\infty} \vert C_{n,\uparrow,g} \vert^2-\lambda
    \end{pmatrix}=0.
    \label{eq95}
    \end{align}
    Solving Eq.~\ref{eq95}, the eigenvalues are obtained as
    \begin{align}
        \lambda_1=\frac{1+\delta}{2},\quad\lambda_2=\frac{1-\delta}{2},
    \end{align}
    where $ \delta = \sqrt{X^2 - 4Y^2}$ with $Y = \sum_{n=1}^{\infty}  \cos(\Omega_n t) \sin(\Omega_{n-1} t)\cdot e^{-|\alpha|^2 } \frac{\alpha^{2n-1}}{{(n-1)!}\sqrt{n}}$.
    Similarly, the reduced density matrix of any single qubit can be expressed as
    \begin{align}
    \rho_a(t) = \mathrm{Tr}_{a}\left( \rho_q(t) \right)
    = \left( \sum_{n=0}^{\infty} \vert C_{n+1,\downarrow,e} \vert^2  \right) \ket{e}\bra{e}+ \left( \sum_{n=0}^{\infty} \vert C_{n,\uparrow,g } \vert^2  \right) \ket{g}\bra{g}
    \end{align}
    By solving the characteristic equation, the eigenvalues of the reduced density matrix $\rho_a(t)$ are obtained as
    \begin{align}
        \lambda_3= P_\uparrow,\quad\lambda_4=P_e,
    \end{align}
    The von Neumann entropy is defined as $E = -\mathrm{Tr}(\rho \log_2 \rho)$. For the entanglement between the magnon and the rest of the system, the von Neumann entropy is written as
    \begin{align}
    E_{\mathrm{M|AS}} &=-\mathrm{Tr}(\rho_q \log_2 \rho_q)= -\frac{1+\delta}{2} \log_2\left( \frac{1+\delta}{2} \right) - \frac{1-\delta}{2} \log_2\left( \frac{1-\delta}{2} \right).
    \end{align}
    The von Neumann entropy $E_{\mathrm{A|MS}}$ $(E_{\mathrm{S|MA}})$ is expressed as
    \begin{equation}
     E_{\mathrm{A|MS}}~(E_{\mathrm{S|MA}}) = -\mathrm{Tr}(\rho_a \log_2 \rho_a) = -P_\uparrow \log_2 P_\uparrow - P_e \log_2 P_e ,
    \end{equation}
    which is used to measure the entanglement between any qubit and the rest of the system. As shown in Fig.~\ref{fg7} (b), influenced by the series summation $Y$, $E_{\mathrm{A|MS}}$ gradually decreases to a minimum and then increases. While the variation of $E_{\mathrm{A|MS}}~(E_{\mathrm{S|MA}})$ is consistent with that of the populations and exhibits collapse and revival. Furthermore, it is obvious that the variation of entanglement quantified by calculating von Neumann entropy or concurrence analytically is consistent with that quantified by logarithmic negativity.

    Then we calculate the concurrence to quantify the bipartite entanglement between two qubits~\cite{PhysRevLett.78.5022}. For the density matrix $\rho_q$ of a two-qubit system, the concurrence is defined as:
    \begin{equation}
    C(\rho_q) = \max(0, \sqrt{\delta_1} - \sqrt{\delta_2} - \sqrt{\delta_3} - \sqrt{\delta_4}),
    \end{equation}
    where \(\delta_1\), \(\delta_2\), \(\delta_3\), and \(\delta_4\) are the eigenvalues of the matrix \( R = \rho_q\tilde{\rho}_q \) with the spin-flipped density matrix \(\tilde{\rho}_q = (\sigma_y \otimes \sigma_y) \rho_q^* (\sigma_y \otimes \sigma_y)\). Through calculations, we obtain the concurrence as
    \begin{equation}
        C_\mathrm{A|S}=2Y.
    \end{equation}
    As shown in Fig.~\ref{fg7} (B), consistent with the change of $Y$, the concurrence $C_\mathrm{A|S}$ first increases to its maximum and then decreases in the collapse region.
    \subsection{Entanglement redistribution in the presence of dissipation}
    \begin{figure}[tb]
        \centering
        \includegraphics[width=0.6\linewidth]{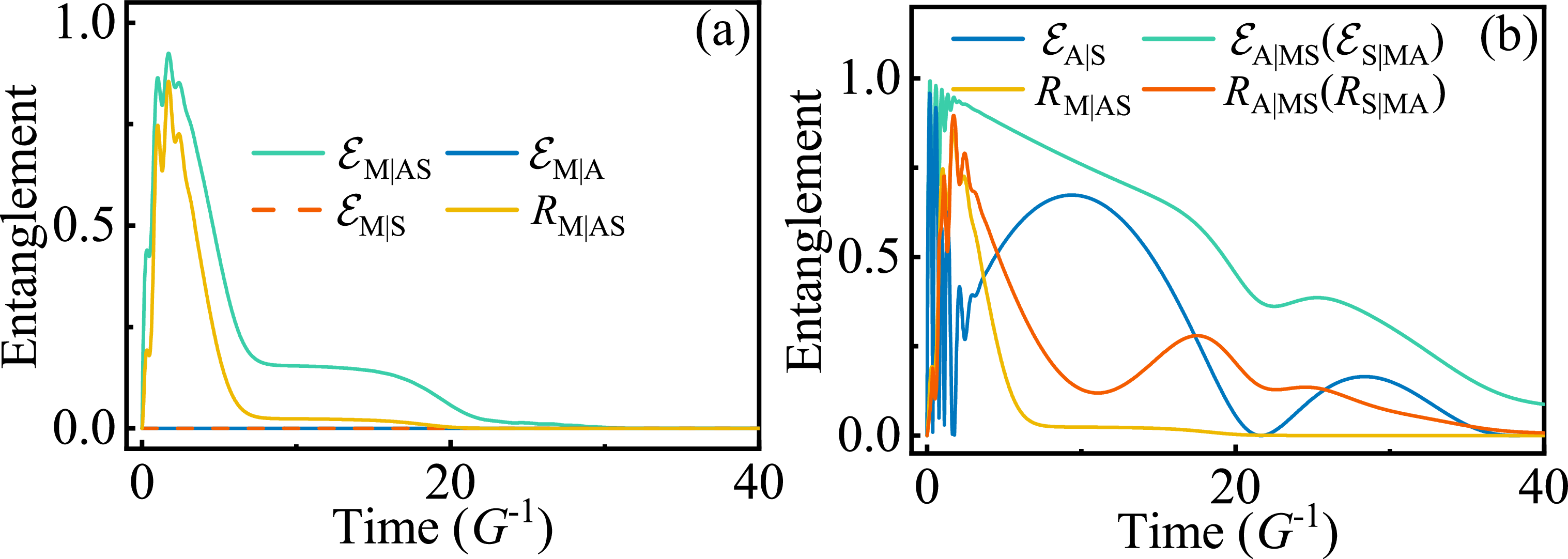}
        \caption{ The evolution of the bipartite entanglement and the residual entanglements when dissipation is considered. The corresponding dissipation rate is chosen as $\kappa_m/2\pi=0.1~\mathrm{MHz}$, $\kappa_a/2\pi=0.01~\mathrm{MHz}$, $\kappa_s/2\pi=0.05~\mathrm{MHz}$, $\gamma_a/2\pi=0.1~\mathrm{MHz}$ and $\gamma_s/2\pi=0.05~\mathrm{MHz}$.}
        \label{f8}
    \end{figure}
    Next, we discuss the evolution of entanglement in the presence of dissipation. The dynamical evolution of the system can be described by the master equation expressed as
    \begin{align}
    \frac{\partial}{\partial t}\rho =&-i\left[ H_\text{eff},\rho \right] +\sum_k\gamma_k \mathcal{L}_{o_k}\left[ \rho \right],
    \end{align}

    where $\rho$ is the density matrix of the system and $\mathcal{L}_{o_k}\left[ \rho \right] ={o_k}\rho {o_k}^\dagger-({o_k}^{\dagger}{o_k}\rho -\rho {o_k}^{\dagger}{o_k})/2$ is the Lindblad superoperators for a given operator ${o_k}$ (${o_k}=m,\sigma ^-_a,\sigma ^-_s,\sigma ^z_a,\sigma ^z_s$) with dissipation rate $\gamma_k$ (${\gamma_k}=\kappa_m,\kappa_a,\kappa_s,\gamma_a,\gamma_s$). Then we can use the QUTIP package in PYTHON to numerically simulate the evolution of entanglement with the dissipation~\cite{johansson2012qutip}. As shown in Fig.~\ref{f8} (a), the logarithmic negativity $\mathcal{E}_\mathrm{M|A}$ and $\mathcal{E}_\mathrm{M|S}$ are always zero, which implies that there is still no bipartite entanglement between the magnon and any qubit when dissipation is taken into account. Furthermore, due to the influence of dissipation, both $\mathcal{E}_\mathrm{M|AS}$ and the residual entanglement $R_\mathrm{M|AS}$ gradually decrease to zero, and no oscillatory recovery of entanglement occurs.
    For the bipartition between any qubit and the rest of the system, we can obtain the relation
    \begin{equation}
        R_\mathrm{A|MS}~(R_\mathrm{S|MA})+\mathcal{E}_\mathrm{A|S}^2=\mathcal{E}_\mathrm{A|MS}^2~(\mathcal{E}_\mathrm{S|MA}^2),
        \label{eq109}
    \end{equation}
    which is the same as the equation without dissipation, since $\mathcal{E}_\mathrm{M|A}~(\mathcal{E}_\mathrm{M|S}=0)$ remains invariant. As shown in Fig.~\ref{f8} (b), $\mathcal{E}_\mathrm{A|MS}~(\mathcal{E}_\mathrm{S|MA})$ is no longer conserved but instead decreases continuously due to the influence of dissipation. However, the residual entanglement $R_\mathrm{A|MS}~(R_\mathrm{S|MA})$ still decreases as $\mathcal{E}_\mathrm{A|S}$ increases and vice versa. Since total entanglement $\mathcal{E}_\mathrm{A|MS}~(\mathcal{E}_\mathrm{S|MA})$ continues to decrease, the increased bipartite entanglement between two qubits $\mathcal{E}_\mathrm{A|C}$ must come at the expense of the reduction in residual entanglement $R_\mathrm{A|MS}~(R_\mathrm{S|MA})$ according to Eq.~\ref{eq109}. So the redistribution of the entanglement from the residual entanglement $R_{\mathrm{A|MS}}$ ($R_{\mathrm{S|MA}}$) to the bipartite entanglement between the two qubits and vice versa still persists when dissipation is taken into account.

    \subsection{Comparison with the Jaynes-Cummings model}
    In this section, we investigate the collapse-revival phenomena and the evolution of entanglement based on JC model in the system consist of a bosonic mode and two qubits. The Hamiltonian of the system can be expressed as
    \begin{equation}
    \hat{H}_\mathrm{JC}=\hbar\omega_m\hat{m}^{\dagger}\hat{m}+\frac{\hbar}{2}\omega_a\hat{\sigma}_a^z+\frac{\hbar}{2}\omega_s\hat{\sigma}_s^z+g(\hat{m}^\dagger\hat{\sigma}^-_a+\hat{m} \hat{\sigma}^+_a) +g(\hat{m}^\dagger\hat{\sigma}^-_s+\hat{m} \hat{\sigma}^+_s),
    \end{equation}
    \begin{figure}[htbp]
        \centering
        \includegraphics[width=0.9\linewidth]{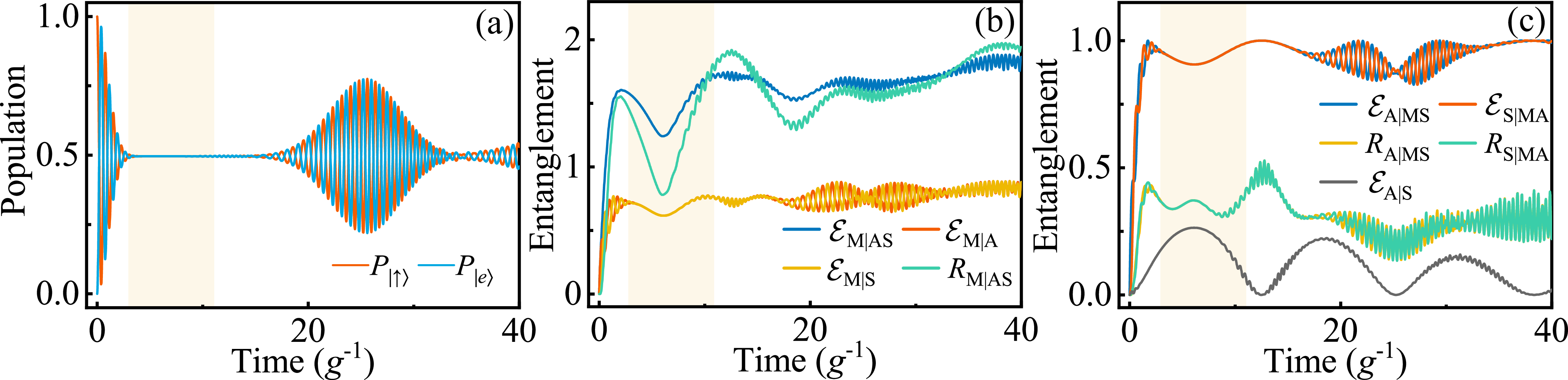}
        \caption{(a) Simulated evolution of populations based on JC model. (b) and (c) shows the evolution of the bipartite entanglement and the residual entanglement based on JC model.}
        \label{f9}
    \end{figure}
    where the resonance condition is selected as $\omega_m=\omega_a=\omega_s$. The magnon is also prepared in the coherent state and $\ket{\alpha}$ with $\alpha=4$. One qubit is prepared in the excited state $\ket{\uparrow}$, and the other in the ground state $\ket{g}$, which is the same as the initial state under the three-body interaction model. Then we calculate the populations of the ASQ $\ket{\uparrow}$ state and the SCQ $\ket{e}$ state. As shown in Fig.~\ref{f9} (a), the collapsed and revival phenomenon of populations can be observed. However, as time progresses, the populations in this figure soon start to exhibit slight oscillations, which result in the collapse region in the Fig.~\ref{f9} (a) is significantly smaller than that based on three-body interaction. Different from the entanglement evolution based on three-body interaction, the entanglement evolution based on JC model is chaotic and trivial. As shown in Fig.~\ref{f9} (b), $\mathcal{E}_\mathrm{M|A}$ and $\mathcal{E}_\mathrm{M|S}$ exhibits a slight variation within the collapse region: first decreasing and then increasing, while there is always no entanglement between the magnon and any qubit under the three-body interaction. Meanwhile, we can also observe that, different from the results based on three-body interaction, $\mathcal{E}_\mathrm{A|MS}$ and $\mathcal{E}_\mathrm{S|MA}$ do not exhibit the collapse and revival in Fig.~\ref{f9} (c). In conclusion, in the Jaynes-Cummings (JC) model, various types of entanglement persist at all times and undergo slight variations over time, which results in the dynamics of entanglement being chaotic and lacking a clear transition. Therefore, the interesting entanglement redistribution—from the three-body form to the bipartite form and vice versa—can only be achieved through three-body interaction.

%
\end{bibunit}
\thispagestyle{empty}
\end{document}